\documentclass{emulateapj}
\usepackage{natbib}

\newcommand{\kms}{\,km\,s$^{-1}$}\newcommand{\sqcm}{cm$^{-2}$}  
\newcommand{\hst}{\emph{HST}}    \newcommand{\fuse}{\emph{FUSE}}
\newcommand{\tm}{\tablenotemark} \newcommand{\tn}{\tablenotetext}
\newcommand{\qa}{NGC~7469}       \newcommand{\qb}{Mrk~335}
\newcommand{\qc}{HE~0226--4110}  \newcommand{\vlsr}{$v_{\rm LSR}$}  
\newcommand{\hi}{\ion{H}{1}}   \newcommand{\hw}{\ion{H}{2}} 
\newcommand{\os}{\ion{O}{6}}   \newcommand{\cf}{\ion{C}{4}} 
\newcommand{\sif}{\ion{Si}{4}} \newcommand{\nf}{\ion{N}{5}}
\newcommand{\ct}{\ion{C}{3}}   \newcommand{\siw}{\ion{Si}{2}}
\newcommand{\cw}{\ion{C}{2}}   \newcommand{\oi}{\ion{O}{1}}
\newcommand{\few}{\ion{Fe}{2}} \newcommand{\alw}{\ion{Al}{2}}
\newcommand{\sit}{\ion{Si}{3}} \newcommand{\sw}{\ion{S}{2}}
\newcommand{\no}{\ion{N}{1}}   \newcommand{\fet}{\ion{Fe}{3}} 
\newcommand{\caw}{\ion{Ca}{2}} \newcommand{\nao}{\ion{Na}{1}} 
\newcommand{\niw}{\ion{Ni}{2}} 
\newcommand{\aro}{\ion{Ar}{1}} 
\newcommand{\dgr}{$\degr$}

\begin{document}

\title{Exploring the Origin and Fate of the Magellanic Stream with
Ultraviolet and Optical Absorption\footnotemark[1]}
  \footnotetext[1]{Based on observations from the NASA-CNES-CSA {\it Far
  Ultraviolet Spectroscopic Explorer} mission (program P101), operated
  by Johns Hopkins University, supported by NASA contract NAS 5-32985,
  from the NASA/ESA {\it Hubble Space Telescope} (program 9802),
  obtained at the Space Telescope Science Institute, which is operated
  by the Association of Universities for Research in Astronomy, Inc., under
  NASA contract NAS 5-26555, and from the Ultraviolet and Visual Echelle
  Spectrograph (UVES) on the Very Large Telescope (VLT) Unit 2
  (Kueyen) at Paranal, Chile, operated by the European Southern
  Observatory, under program ID 081.D-0697(A).}

\author{Andrew J. Fox$^2$, Bart  P. Wakker$^3$, Jonathan V. Smoker$^2$,
  Philipp Richter$^4$, Blair D. Savage$^2$, \& Kenneth R. Sembach$^5$}
\affil{$^2$ European Southern Observatory, Alonso de C\'ordova
  3107, Casilla 19001, Vitacura, Santiago, Chile; afox@eso.org\\
$^3$ Department of Astronomy, University of
  Wisconsin--Madison, 475 North Charter St., Madison, WI 53706\\
$^4$ Institut f\"ur Physik und Astronomie, Universit\"at Potsdam, Haus
  28, Karl-Liebknecht-Str. 24/25, 14476, Potsdam, Germany\\
$^5$ Space Telescope Science Institute, 3700 San Martin Drive,
  Baltimore, MD 21218}

\shorttitle{Fox et al.}
\shortauthors{Magellanic Stream in Absorption} 

\begin{abstract}
We present an analysis of ionization and metal enrichment in the
Magellanic Stream (MS), the nearest gaseous tidal stream, using
\emph{HST}/STIS and \emph{FUSE} ultraviolet spectroscopy of two
background AGN. The targets are NGC~7469, lying directly behind the MS
with log\,$N$(\hi)$_{\rm MS}$=18.63$\pm$0.03(stat)$\pm$0.08(syst),
and Mrk~335, lying 24.7\dgr\ away with 
log\,$N$(\hi)$_{\rm MS}$=16.67$\pm$0.05.
For NGC~7469, we include optical spectroscopy from VLT/UVES.
In both sightlines the MS is detected in 
low-ion (\oi, \cw, \ct, \siw, \sit, \alw, \caw) and
high-ion (\os, \cf, \sif) absorption.
Toward NGC~7469, we measure a MS oxygen abundance
[O/H]$_{\rm MS}$=[\oi/\hi]=$-$1.00$\pm$0.05(stat)$\pm$0.08(syst), 
supporting the view that the Stream originates in the SMC rather than
the LMC.
We use CLOUDY to model the low-ion phase of the Stream as a
photoionized plasma using the observed \sit/\siw\ and \ct/\cw\ ratios.
Toward Mrk~335 this yields an ionization
parameter log\,$U$ between $-$3.45 and $-$3.15, 
a gas density log\,($n_{\rm H}$/cm$^{-3}$) between $-$2.51 and $-$2.21, 
and a hydrogen ionization fraction of 98.9--99.5\%. 
Toward NGC~7469 we derive sub-solar abundance ratios for [Si/O], [Fe/O],
and [Al/O], indicating the presence of dust in the MS.
The high-ion column densities are too large to be explained by
photoionization, but also cannot be explained by a single-temperature
collisional-ionization model (equilibrium or non-equilibrium).
This suggests the high-ion plasma is multi-phase, with a \sif\ region,
a hotter \os\ region, and \cf\ potentially contributing to each.
Summing over the low-ion and high-ion phases, we derive conservative
lower limits on the ratio $N$(total \hw)/$N$(\hi) of
$\ga$19 toward NGC~7469 and $\ga$330 toward Mrk~335, 
showing that along these two directions the vast majority of the
Stream has been ionized. The presence of warm-hot plasma 
together with the small-scale
structure observed at 21\,cm provides evidence for an evaporative 
interaction with the hot Galactic corona. This scenario, predicted by
hydrodynamical simulations, suggests that the fate of the MS will be to
replenish the Galactic corona with new plasma, rather than to bring
neutral fuel to the disk. 
\end{abstract}

\section{Introduction}
Gaseous inflow is a key process in galaxy evolution,
serving to regulate chemical abundances and provide fuel for star
formation. Without inflow, galactic chemodynamical models cannot
reproduce observed stellar abundance patterns \citep{Ma01}.
In the case of the Milky Way, an average inflow rate of
$\sim$1\,$M_\odot$\,yr$^{-1}$ over the last 10\,Gyr, and a present-day 
inflow rate of $\sim$0.4\,$M_\odot$\,yr$^{-1}$, is needed to solve the G-dwarf
problem, the anomalously narrow abundance distribution of nearby 
solar-type stars \citep{PP75, Ch08}. 
Observationally, this infalling circumgalactic gas
is seen in the form of high-velocity clouds (HVCs), defined
as interstellar clouds having velocities incompatible with co-rotation
with the disk  
\citep[in practice meaning $|v_{\rm  LSR}|\!>\!100$\kms;][]{WW97, Ri06}.
HVCs have been heralded as the solution to the G-dwarf problem
\citep{Wa99}, since the mass inflow rate of neutral gas in 
HVCs is $\sim$0.1--0.4\,$M_\odot$\,yr$^{-1}$ \citep{Wa08, Pe08}.
However, the role of ionization is not fully understood,
and there are indications of a significant mass inflow rate in ionized gas
\citep{BH07, Sh09}, and for a large population of low $N$(\hi) HVC fragments
across the entire sky \citep{Se03, Co05, Co09, Fo06, Ri09}.

The Magellanic Stream (MS) is a large and well-resolved
HVC, and the nearest gaseous tidal stream.
A broad filament of 21\,cm-emitting gas extending for
$\approx\!100$\dgr\ on either side of
the South Galactic Pole \citep{WW72, Ma74, Ma77, Pu03a, BT04, Br05}
and containing $\sim\!10^8$\,M$_\odot$ of neutral gas,
the MS is thought to have been stripped out of
the Magellanic Clouds by either ram pressure or tidal forces.
This origin is supported by \hst\ measurements of the MS metallicity 
\citep[$Z_{\rm MS}$=0.2--0.4 solar;][]{Lu94, Lu98, Gi00, Se01},
and many orbital simulations \citep{MD94, Li95, GN96, Ma05, Co06},
though it is unclear whether the LMC or the SMC is the primary source
of the Stream's material \citep{Ni08}. 

Like many (if not all) HVCs, the MS is a multi-phase structure,
containing regions of molecular gas seen in H$_2$ absorption \citep{Ri01},
cold neutral \citep{Be06,Ma09} and warm neutral \citep{Br05,KH06} gas seen
in 21\,cm emission, warm-ionized plasma seen in H$\alpha$ emission 
\citep[$T\!\sim\!10^4$\,K;][]{WW96, Pu03b} and UV absorption \citep{Lu94}, and 
highly-ionized plasma seen in \os\ absorption 
\citep[$T\!\sim\!10^{5-6}$\,K;][]{Se03, Fo05a}.
Many UV studies \citep{Se95,Se99,Co04,Co05,Fo04,Fo05a,Ga05,Ze08}
have led to the consensus that the high ions (\os, \cf, \sif) in HVCs
trace the boundary layers between the warm cloud cores (seen in \hi\
and the low ions) and a surrounding hot medium. Various theoretical
models have been proposed to explain such boundary layers,
including conductive interfaces \citep{BH87, Bo90, GS10}, 
turbulent mixing layers generated by Kelvin-Helmholtz instabilities  
\citep{Sl93, Es06}, and shock ionization \citep{DS96, IS04}, which
all relate to the interaction between HVCs and the hot
coronal plasma \citep{FF94, Mu00, QM01, Ko01, ML04}. 

Recent simulations have examined the lifetime of HVCs against disruption by
these cloud/corona interactions.
\citet{BH09} find HVC lifetimes against break up by
thermal instability of 60--180\,Myr, which for a typical HVC
velocity of 200\kms\ corresponds to distances of 12--36\,kpc.
In the simulations of \citet{HP09}, 
infalling HVCs with \hi\ masses $\la\!10^{4.5}$\,M$_\odot$ become fully
ionized by Kelvin-Helmholtz instabilities within $\approx$10\,kpc. 
Although the total 21\,cm mass in the MS is several orders of
magnitude larger than this \citep{Br05}, the Stream is known to be
fragmented into smaller clumps \citep{St02, St08, WK08, Ni09}, 
suggesting that a disruptive interaction with the ambient hot corona
is already underway.

Understanding ionization in HVCs is crucial to assess their role in
Galactic evolution \citep{MB04}. If an HVC becomes fully ionized
before reaching the disk, it will not provide fuel for future
star formation unless it stays bound, cools, and
recombines within the inflow timescale ($\sim$few$\times10^8$\,yr).
If instead a cloud  ``evaporates'' before reaching the
disk\footnotemark[6]\footnotetext[6]{By ``evaporation'' we refer to
  the process(es) by which gas is stripped from an HVC and becomes
  incorporated into the surrounding hot plasma; this is different
  from ``ionization'', since a cloud can be ionized
  without being evaporated.} 
it will simply serve to replenish the hot corona with new material.
Determining the ionization level empirically in a nearby massive HVC
(the MS) is the focus of this paper. We also demonstrate the ability of
UV absorption lines to probe and characterize low column density
regions of the MS. 

UV studies of the MS have been hampered by a shortage
of suitable UV-bright background targets. 
MS absorption detections have been reported toward
Fairall~9 behind the MS proper \citep{So81, Lu94, Gi00, Ri01}, 
NGC~3783 behind the leading arm of the Stream \citep{Lu98, Se01},
III~Zw~2 and \qa\ \citep[both near the outer 21\,cm contour of the
  MS;][]{Gi00}, and \qc\ \citep[11\dgr\ away from the 21\,cm
emission;][]{Fo05a}.
In addition, high-velocity absorption reported toward NGC~1705 
\citep{Sa98} may trace an outer fragment of the MS. 
Other targets are available behind the Magellanic Bridge connecting
the LMC and SMC \citep{Le01, Le08, Sm05, Mi09}, but this is usually
considered separately from the MS. Here we present new UV observations
of two sight lines in the vicinity of the MS:
(1) \qa, also known as QSO~B2300+086 and Mrk~1514, a Seyfert~1 Galaxy
at $l,b$=83.10\dgr,$-$45.47\dgr\ lying behind the MS, but
close to the outer 21\,cm contour, with a MS 
log\,$N$(\hi)$\approx$18.63 (see \S4.1), and 
(2) \qb, also known as QSO~B0003+199, a Seyfert~1 Galaxy 
at $l,b$=108.76\dgr,$-$41.42\dgr\ lying 24.9\dgr\ 
away from \qa\ in a direction with no MS 21\,cm emission detected,
even in the most sensitive \hi\ observations 
\citep[which reach log\,$N$(\hi)$\approx$17.5;][]{BT04}. 
The location of these two targets relative to the 21\,cm 
emission from the MS is shown in Figure 1.

\begin{figure}
\includegraphics[height=10cm,angle=270]{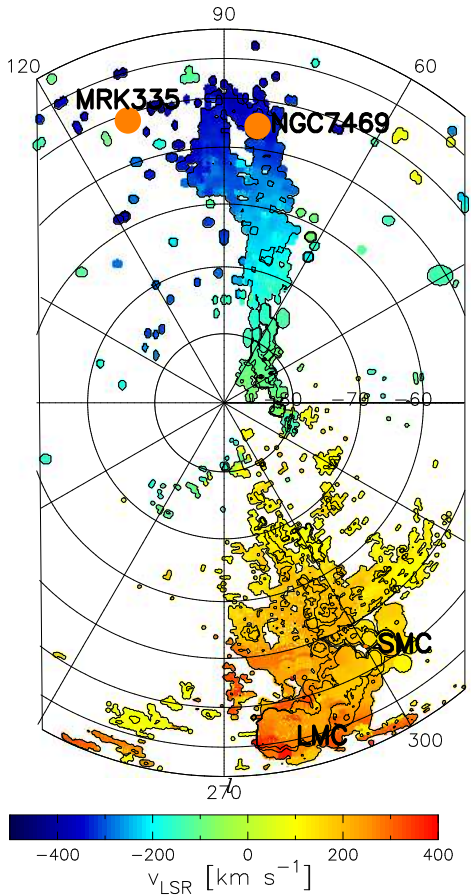}
\caption{\hi\ map of the Magellanic Stream color-coded by
velocity and centered on the South Galactic Pole, using 21\,cm data from
\citet{HW88} and \citet{Mr00} with a sensitivity of 
log\,$N$(\hi)$\approx$18.30.
The positions of our two sight lines are marked.}
\end{figure}

This paper is organized as follows. In \S2 we discuss the data origin
and handling. In \S3 we discuss the spectroscopic measurements,
and in \S4 an overview of the MS absorption is presented. 
In \S5 we derive the metallicity of the MS.
In \S6 and \S7 we present photo-ionization models (for the low ions) and
collisional ionization models (for the high ions), respectively. In
\S8 we discuss the relative amounts of neutral and ionized gas in the MS, 
and the implications for the fate of the Stream. 
Our results are summarized in \S9.

\section{Observations and Data Handling}
\qa\ and \qb\ were observed using the E140M echelle grating of the
Space Telescope Imaging Spectrograph \citep[STIS;][]{Wo98, Ki98, Bo98},
which provides high-resolution (7.0\kms\ FWHM) spectra in the range
1150--1700\,\AA. These observations were taken in 2004 during \hst\ Cycle 12
under proposal ID 9802, using the 0.2$\times$0.06\arcsec\ aperture
and the FUV-MAMA detectors in ACCUM mode for
total exposure times of 22\,810\,s (\qa) and 17\,105\,s (\qb). 
The calibrated STIS data were retrieved from the \hst\ MAST archive,
having been processed with the CalSTIS pipeline \citep[v2.19;][]{Dr07}.
The relative wavelength calibration of STIS is accurate to
0.1--0.4 pixels, with an absolute calibration accurate to 0.2--0.5
pixels, or 0.6--1.6\kms\ \citep{KQ07}. These data were rebinned to
three pixels (i.e. to 9.6\kms\ bins) for display purposes, but all 
measurements were made on the unbinned data.

\qa\ and \qb\ were observed in 2000 and 2001 with the 
{\it Far Ultraviolet Spectroscopic Explorer} (\fuse) satellite
\citep{Mo00, Sh00}, providing medium-resolution 
spectra ($\approx$20--25\kms\ FWHM) in the far-ultraviolet range
912--1187\,\AA\ in two channels (using LiF and SiC coatings).
These observations were taken during \fuse\ Cycle 1
under proposal ID P101, and used the LWRS aperture for
total exposure times of 37\,630\,s (\qa) and 99\,284\,s (\qb). 
The raw \fuse\ data were processed with version 2.4.0 of the CalFUSE
pipeline \citep{Di07}, following the data reduction procedures
described in \citet{Wa03}, which involve measuring the centroids of various
atomic (\aro, \oi, \siw) and molecular (H$_2$) interstellar
lines in each segment of each exposure, applying 
offsets to bring them onto the same (LSR) velocity scale as the STIS
observations, and then co-adding the aligned spectra.
The reduced \fuse\ data have an absolute
wavelength calibration accurate to 5--10\kms, and were rebinned to
five pixels (i.e. to 10.0\kms\ bins) for display purposes.

Fortuitously, a Type Ia supernova (SN 2008ec) was discovered in \qa\ 
on 14 July 2008 at position RA=23$^{\rm h}$03$^{\rm m}$16.56$^{\rm s}$,
Dec=+08\dgr52\arcmin19.8\arcsec\
\citep[J2000;][]{Re08}, 6.5\arcsec\ away from the center of the 
\fuse\ pointing. 
This event was observed with the UVES high-resolution optical
spectrograph \citep{De00} mounted on the 8.2\,m Kueyen unit 
of the Very Large Telescope (VLT) at Cerro Paranal, Chile,
under ESO program ID 081.D-0697(A). Three epochs of
observation were taken on 26 July 2008, 9 August 2008, and 26 August 2008
using the Dichroic 1 390+564 setting, a 1.0\arcsec\ slit, and
1$\times$1 binning, all at low airmass (1.2--1.3).
These optical data allow us to search for 
\caw\ $\lambda\lambda$ 3934.777, 3969.591 and 
\nao\ $\lambda\lambda$ 5891.583, 5897.558 absorption
in the MS. The spectra were extracted from the ESO archive and reduced
using the UVES pipeline \citep{Ba00} in the ESOREX environment.
Because of fading of the supernova, the S/N of 
the second- and third-epoch spectra is lower, so only some of the individual
exposures were combined (using median combining) to form our final spectrum. 
In the blue arm (390\,nm setting) we used only the first-epoch exposures
(6$\times$1\,800\,s). In the red arm (564\,nm setting), 
we used 3$\times$1\,800\,s from the first epoch and a total of
20\,400\,s from the second epoch. 
These UVES data have a spectral resolution of $\sim$40\,000 or 7.5\kms (FWHM),
and were rebinned to two pixels (2.2\kms\ bins) for display and
measurement.

We also make use of 21\,cm \hi\ emission-line spectra, to constrain $N$(\hi)
in the MS along each line of sight. Spectra from several radio
telescopes are available, including the Green Bank 140\,foot
telescope, Green Bank 100\,m Telescope (GBT), the Effelsberg 100\,m
telescope, and the Leiden-Argentine-Bonn Survey \citep{Ka05}. 
The choice of dataset used for each sightline is discussed in \S4.1 and 4.2.

All wavelengths and velocities in this paper are presented relative to
the Local Standard of Rest (LSR), as defined by \citet{KL86}; 
the corrections from the heliocentric frame
are \vlsr=$v_{\rm HEL}$+6.0\kms\ for \qa\ and 
\vlsr=$v_{\rm HEL}$+3.3\kms\ for \qb.
For all absorption-line data (STIS, \fuse, and UVES), 
continua were fitted locally around
each line of interest, using polynomial fits to spectral regions
several hundred \kms\ wide, and the data were normalized by the continuum.

\section{Measurement of Absorption}
High-negative-velocity ($v_{\rm LSR}\!<\!-100$\kms) absorption is 
present in the spectra of both \qa\ and \qb\ in a range of low-ion and
high-ion lines. Two techniques were used to measure the HV absorption:
Voigt-profile fitting and apparent optical depth (AOD) measurements. 
These measurements are described in the next two sub-sections,
yielding the results presented in Table 1 (for \qa) and Table 2 (for \qb). Rest
wavelengths and oscillator strengths were taken from \citet{Mo03, Mo04}. 
Signal-to-noise measurements were made locally near each line, by
measuring the r.m.s. dispersion of the data around the continuum in
regions 100\kms\ wide.

\begin{deluxetable*}{lcccc cccc}
\tablewidth{0pt}
\tabletypesize{\scriptsize}
\tablecaption{Measurements of High-Velocity Absorption toward NGC 7469}
\tablehead{& & \multicolumn{3}{c}{\underline{~~~~~~~~~~~~~~AOD Results\tm{c}~~~~~~~~~~~~~~}} & \multicolumn{4}{c}{\underline{~~~~~~~~~~~~~~~~~~~~~VPFIT Results\tm{d}~~~~~~~~~~~~~~~~~~~}}\\ Line & S/N\tm{a} & Component\tm{b} & $W_\lambda$ & log\,$N_a$ & $v_0$ & $b$ & log\,$N$ & log\,$N$(tot)\tm{e}\\ & & & (m\AA) & ($N_a$ in \sqcm) & (km\,s$^{-1}$) & (km\,s$^{-1}$) & ($N$ in \sqcm) & ($N$ in \sqcm)}
\startdata
  \cw\ $\lambda$1334.532\tm{g} &       16 &            MS & $>$292 &             $>$14.49 &              $-$360$\pm$1 & 23$\pm$4 & 14.01$\pm$0.05 & 14.45$\pm$0.04 \\
\nodata & \nodata & \nodata & \nodata & \nodata & $-$330$\pm$1 & 13$\pm$2 & 14.19$\pm$0.05 & \nodata \\
                       \nodata &  \nodata &       HVC1 &  31$\pm$6 &       13.25$\pm$0.07 &                     $-$251$\pm$2 & 11$\pm$3 & 13.25$\pm$0.08 & \nodata \\
                       \nodata &  \nodata &           HVC2 & $<$44 &             $<$13.34 &             $-$184$\pm$2 &  5$\pm$4 & [12.90$\pm$0.13]\tm{f} & \nodata \\
 \siw\ $\lambda$1260.422\tm{h} &       30 &            MS & $>$251 &             $>$13.51 &              $-$368$\pm$2 & 13$\pm$4 & 13.16$\pm$0.08 & 13.60$\pm$0.04 \\
\nodata & \nodata & \nodata & \nodata & \nodata & $-$335$\pm$1 & 10$\pm$2 & 13.41$\pm$0.05 & \nodata \\
       \siw\ $\lambda$1193.290 &       10 &        MS & 176$\pm$17 &       13.64$\pm$0.05 &                                  \nodata & \nodata & \nodata & \nodata \\
       \siw\ $\lambda$1526.707 &       12 &        MS &  73$\pm$16 &       13.56$\pm$0.08 &                                  \nodata & \nodata & \nodata & \nodata \\
       \siw\ $\lambda$1304.370 &       21 &        MS &  47$\pm$10 &       13.59$\pm$0.08 &                                  \nodata & \nodata & \nodata & \nodata \\
       \siw\ $\lambda$1190.416 &        9 &        MS & 152$\pm$17 &       13.84$\pm$0.05 &                                  \nodata & \nodata & \nodata & \nodata \\
       \sit\ $\lambda$1206.500 &       14 &            MS & $>$404 &             $>$13.82 &          \nodata\tm{i} & \nodata\tm{i} & \nodata\tm{i} & \nodata\tm{i} \\
                       \nodata &  \nodata &          HVC1 & $>$ 55 &             $>$12.81 &                     $-$248$\pm$1 &  4$\pm$2 & 12.87$\pm$0.45 & \nodata \\
                       \nodata &  \nodata &      HVC2 &  52$\pm$13 &       12.56$\pm$0.08 &                     $-$178$\pm$2 & 12$\pm$3 & 12.53$\pm$0.08 & \nodata \\
        \oi\ $\lambda$1302.168 &       23 &        MS & 107$\pm$10 &       14.32$\pm$0.04 &              $-$364$\pm$1 &  6$\pm$6 & 13.69$\pm$0.09 & 14.29$\pm$0.04 \\
\nodata & \nodata & \nodata & \nodata & \nodata & $-$337$\pm$1 &  8$\pm$2 & 14.17$\pm$0.05 & \nodata \\
                       \nodata &  \nodata &           HVC1 & $<$24 &             $<$13.51 &                                  \nodata & \nodata & \nodata & \nodata \\
                       \nodata &  \nodata &           HVC2 & $<$31 &             $<$13.63 &                                  \nodata & \nodata & \nodata & \nodata \\
        \no\ $\lambda$1199.550 &       14 &             MS & $<$57 &             $<$13.54 &                                  \nodata & \nodata & \nodata & \nodata \\
                       \nodata &  \nodata &           HVC1 & $<$47 &             $<$13.45 &                                  \nodata & \nodata & \nodata & \nodata \\
                       \nodata &  \nodata &           HVC2 & $<$53 &             $<$13.50 &                                  \nodata & \nodata & \nodata & \nodata \\
       \alw\ $\lambda$1670.787 &       18 &        MS & 110$\pm$20 &       12.46$\pm$0.07 &             $-$364$\pm$14 & 20$\pm$5 & 12.02$\pm$0.19 & 12.45$\pm$0.12 \\
\nodata & \nodata & \nodata & \nodata & \nodata & $-$327$\pm$4 & 15$\pm$6 & 12.25$\pm$0.14 & \nodata \\
                       \nodata &  \nodata &           HVC1 & $<$57 &             $<$12.09 &                                  \nodata & \nodata & \nodata & \nodata \\
                       \nodata &  \nodata &           HVC2 & $<$93 &             $<$12.30 &                                  \nodata & \nodata & \nodata & \nodata \\
       \few\ $\lambda$1144.939 &       29 &             MS & $<$41 &             $<$13.51 &      $-$340$\pm$1 & 12$\pm$4 & [13.19$\pm$0.07]\tm{f} & 13.19$\pm$0.07 \\
                       \nodata &  \nodata &           HVC1 & $<$26 &             $<$13.31 &                                  \nodata & \nodata & \nodata & \nodata \\
                       \nodata &  \nodata &           HVC2 & $<$31 &             $<$13.39 &                                  \nodata & \nodata & \nodata & \nodata \\
       \few\ $\lambda$1608.451 &       10 &             MS & $<$62 &             $<$13.70 &                                  \nodata & \nodata & \nodata & \nodata \\
       \fet\ $\lambda$1122.526 &       24 &             MS & $<$53 &             $<$13.94 &      $-$321$\pm$2 & 16$\pm$4 & [13.63$\pm$0.05]\tm{f} & 13.63$\pm$0.05 \\
                       \nodata &  \nodata &           HVC1 & $<$30 &             $<$13.70 &                                  \nodata & \nodata & \nodata & \nodata \\
        \sw\ $\lambda$1259.519 &       26 &             MS & $<$20 &             $<$13.94 &                                  \nodata & \nodata & \nodata & \nodata \\
                       \nodata &  \nodata &           HVC1 & $<$12 &             $<$13.71 &                                  \nodata & \nodata & \nodata & \nodata \\
       \caw\ $\lambda$3934.777 &       86 &         MS &  23$\pm$3 &       11.43$\pm$0.05 &              $-$334$\pm$1 &  7$\pm$2 & 11.40$\pm$0.05 & 11.40$\pm$0.05 \\
                       \nodata &  \nodata &           HVC1 & $<$11 &             $<$11.08 &                                  \nodata & \nodata & \nodata & \nodata \\
                       \nodata &  \nodata &           HVC2 & $<$11 &             $<$11.10 &                                  \nodata & \nodata & \nodata & \nodata \\
       \caw\ $\lambda$3969.591 &      100 &         MS &  14$\pm$3 &       11.49$\pm$0.08 &                                  \nodata & \nodata & \nodata & \nodata \\
       \nao\ $\lambda$5891.583 &       61 &             MS & $<$25 &             $<$11.12 &                                  \nodata & \nodata & \nodata & \nodata \\
                       \nodata &  \nodata &           HVC1 & $<$11 &             $<$10.74 &                                  \nodata & \nodata & \nodata & \nodata \\
                       \nodata &  \nodata &           HVC2 & $<$15 &             $<$10.88 &                                  \nodata & \nodata & \nodata & \nodata \\
        \cf\ $\lambda$1548.205 &       15 &         MS & 297$\pm$9 &       14.02$\pm$0.02 &              $-$329$\pm$2 & 47$\pm$3 & 14.02$\pm$0.03 & 14.24$\pm$0.03 \\
\nodata & \nodata & \nodata & \nodata & \nodata & $-$294$\pm$3 &  4$\pm$4 & 12.69$\pm$0.28 & \nodata \\
                       \nodata &  \nodata &       HVC1 &  70$\pm$6 &       13.36$\pm$0.04 &                     $-$249$\pm$1 &  8$\pm$2 & 13.21$\pm$0.08 & \nodata \\
                       \nodata &  \nodata &       HVC2 & 142$\pm$9 &       13.75$\pm$0.03 &                     $-$182$\pm$1 & 12$\pm$2 & 13.66$\pm$0.04 & \nodata \\
        \cf\ $\lambda$1550.778 &       23 &        MS & 189$\pm$10 &       14.06$\pm$0.03 &                                  \nodata & \nodata & \nodata & \nodata \\
                       \nodata &  \nodata &           HVC1 & $<$53 &             $<$13.41 &                                  \nodata & \nodata & \nodata & \nodata \\
                       \nodata &  \nodata &      HVC2 &  76$\pm$13 &       13.70$\pm$0.07 &                                  \nodata & \nodata & \nodata & \nodata \\
       \sif\ $\lambda$1393.755 &       13 &        MS & 150$\pm$11 &       13.33$\pm$0.03 &              $-$327$\pm$2 & 28$\pm$3 & 13.27$\pm$0.04 & 13.43$\pm$0.03 \\
                       \nodata &  \nodata &       HVC1 &  28$\pm$7 &       12.57$\pm$0.10 &                     $-$250$\pm$2 &  9$\pm$3 & 12.56$\pm$0.10 & \nodata \\
                       \nodata &  \nodata &      HVC2 &  39$\pm$10 &       12.76$\pm$0.09 &                     $-$183$\pm$1 &  6$\pm$3 & 12.69$\pm$0.07 & \nodata \\
       \sif\ $\lambda$1402.770 &       14 &        MS &  61$\pm$10 &       13.20$\pm$0.06 &                                  \nodata & \nodata & \nodata & \nodata \\
                       \nodata &  \nodata &           HVC1 & $<$20 &             $<$12.64 &                                  \nodata & \nodata & \nodata & \nodata \\
                       \nodata &  \nodata &           HVC2 & $<$26 &             $<$12.76 &                                  \nodata & \nodata & \nodata & \nodata \\
        \nf\ $\lambda$1238.821 &       20 &             MS & $<$13 &             $<$12.78 &                                  \nodata & \nodata & \nodata & \nodata \\
                               &          &           HVC1 & $<$10 &             $<$12.66 &                                  \nodata & \nodata & \nodata & \nodata \\
                               &          &           HVC2 & $<$15 &             $<$12.83 &      $-$177$\pm$3 & 10$\pm$6 & [12.63$\pm$0.18]\tm{f} & 12.63$\pm$0.18 \\
  \os\ $\lambda$1031.926\tm{j} &       24 &        MS & 119$\pm$10 &       14.09$\pm$0.03 &            $-$333$\pm$ 9 & 31$\pm$10 & 13.95$\pm$0.16 & 13.95$\pm$0.16 \\
\enddata
\tn{a}{~Signal-to-noise ratio per resolution element in continuum next to line.\\}
\tn{b}{~AOD measurements were made separately in each HVC component, using the following LSR velocity ranges. MS: $-$400 to $-$265\kms; HVC1: $-$265 to $-$220\kms; HVC2: $-$220 to $-$130\kms. For certain lines/components, the velocity ranges were slightly adjusted to avoid contamination or addition of noise, as follows: \few\ 1144 and \fet\ 1122, MS, $-$350 to $-$265\kms; \few\ 1608, MS, $-$350 to $-$300\kms; \caw\ 3934,3969, MS, $-$360 to $-$300\kms; \nf\ 1238,1242, HVC2, $-$200 to $-$150\kms.\\}
\tn{c}{~Upper limits on $W_\lambda$ and log\,$N_a$ for non-detections (lines not detected at 3$\sigma$ significance) are calculated as $W_\lambda$(actual)+3$\sigma$ (e.g. if $W_\lambda$=20$\pm$10\,m\AA\ we quote $<$50\,m\AA). Lower limits are for saturated lines. Errors on  $W_\lambda$ and log\,$N_a$ include both statistical and continuum placement uncertainties (and 6\,m\AA\ fixed pattern noise for \fuse\ lines). When a component is not detected in multiple lines of a particular ion, only the strongest constraint on log\,$N$ is given.\\}
\tn{d}{~All lines of a given ion were fitted simultaneously, so VPFIT results are presented only once per ion. Two MS components were fitted for \cw, \oi, \siw, \sit, \alw, and \cf.\\}
\tn{e}{~Total VPFIT column density, summed over all fitted HVC components, with errors added in quadrature.\\}
\tn{f}{~VPFIT conducted even though absorption in this component not detected at 3$\sigma$ significance; results uncertain.\\}
\tn{g}{~No measurement of HVC absorption in \cw\ $\lambda$1036.337 made due to contamination with Galactic H$_2$ $\lambda$1035.183 and $\lambda$1035.427.\\}
\tn{h}{~No measurement of \siw\ $\lambda$1260.422 made in HVC1 or HVC2 due to contamination with Galactic \sw\ 1259.519.\\}
\tn{i}{~\sit\ $\lambda$1206.500 saturated in MS component; no reliable fit possible.\\}
\tn{j}{~No measurement of \os\ $\lambda$1031.926 made in HVC1 or HVC2 due to contamination with Galactic H$_2$ $\lambda$1031.193.\\}
\end{deluxetable*}

\begin{deluxetable*}{lcccc cccc}
\tablewidth{0pt}
\tabletypesize{\scriptsize}
\tablecaption{Measurements of High-Velocity Absorption toward Mrk 335}
\tablehead{& & \multicolumn{3}{c}{\underline{~~~~~~~~~~~~~~AOD Results\tm{c}~~~~~~~~~~~~~~}} & \multicolumn{4}{c}{\underline{~~~~~~~~~~~~~~~~~~~~~VPFIT Results\tm{d}~~~~~~~~~~~~~~~~~~~}}\\ Line & S/N\tm{a} & Component\tm{b} & $W_\lambda$ & log\,$N_a$ & $v_0$ & $b$ & log\,$N$ & log\,$N$(tot)\tm{e}\\ & & & (m\AA) & ($N_a$ in \sqcm) & (km\,s$^{-1}$) & (km\,s$^{-1}$) & ($N$ in \sqcm) & ($N$ in \sqcm)}
\startdata
        \hi\ $\lambda$ 923.150 &       14 &            MS & $>$219 &             $>$16.49 &        $-$341$\pm$2 & 58$\pm$2\tm{i} & 16.67$\pm$0.05 & 16.67$\pm$0.05 \\
        \hi\ $\lambda$ 920.963 &       14 &            MS & $>$197 &             $>$16.55 &                                  \nodata & \nodata & \nodata & \nodata \\
  \cw\ $\lambda$1334.532\tm{g} &       15 &      HVC3 &  60$\pm$11 &       13.54$\pm$0.07 &              $-$415$\pm$1 &  8$\pm$2 & 13.18$\pm$0.04 & 13.68$\pm$0.02 \\
                       \nodata &  \nodata &         MS &  51$\pm$7 &       13.51$\pm$0.06 &                     $-$332$\pm$1 & 12$\pm$2 & 13.51$\pm$0.03 & \nodata \\
  \ct\ $\lambda$ 977.020\tm{h} &       11 &      HVC3 & 132$\pm$10 &       13.48$\pm$0.03 &              $-$407$\pm$6 & 31$\pm$5 & 13.30$\pm$0.12 & 14.12$\pm$0.05 \\
                       \nodata &  \nodata &            MS & $>$249 &             $>$14.00 &                    $-$340$\pm$1 & 37$\pm$12 & 13.51$\pm$0.15 & \nodata \\
\nodata & \nodata & \nodata & \nodata & \nodata & $-$303$\pm$1 & 36$\pm$3 & 13.90$\pm$0.04 & \nodata \\
       \siw\ $\lambda$1260.422 &       23 &       HVC3 &  39$\pm$7 &       12.50$\pm$0.07 &              $-$426$\pm$2 & 14$\pm$4 & 12.40$\pm$0.10 & 12.74$\pm$0.06 \\
                       \nodata &  \nodata &         MS &  35$\pm$5 &       12.49$\pm$0.06 &                     $-$339$\pm$1 &  9$\pm$2 & 12.48$\pm$0.08 & \nodata \\
       \siw\ $\lambda$1193.290 &        6 &          HVC3 & $<$108 &             $<$13.23 &                                  \nodata & \nodata & \nodata & \nodata \\
                       \nodata &  \nodata &             MS & $<$61 &             $<$12.98 &                                  \nodata & \nodata & \nodata & \nodata \\
       \sit\ $\lambda$1206.500 &       12 &          HVC3 & $<$105 &             $<$12.69 &      $-$421$\pm$1 &  5$\pm$5 & [12.47$\pm$0.61]\tm{f} & 13.26$\pm$0.46 \\
\nodata & \nodata & \nodata & \nodata & \nodata & $-$389$\pm$2 &  8$\pm$3 & 12.23$\pm$0.15 & \nodata \\
                       \nodata &  \nodata &        MS & 146$\pm$15 &       13.01$\pm$0.06 &                     $-$336$\pm$2 & 18$\pm$4 & 12.74$\pm$0.08 & \nodata \\
\nodata & \nodata & \nodata & \nodata & \nodata & $-$300$\pm$2 &  9$\pm$3 & 12.63$\pm$0.10 & \nodata \\
                       \nodata &  \nodata &           HVC4 & $<$78 &             $<$12.56 &             $-$249$\pm$4 &  3$\pm$3 & [12.56$\pm$1.00]\tm{f} & \nodata \\
        \oi\ $\lambda$1302.168 &       14 &           HVC3 & $<$51 &             $<$13.84 &                                  \nodata & \nodata & \nodata & \nodata \\
                       \nodata &  \nodata &             MS & $<$39 &             $<$13.73 &                                  \nodata & \nodata & \nodata & \nodata \\
        \no\ $\lambda$1199.550 &        7 &           HVC3 & $<$73 &             $<$13.64 &                                  \nodata & \nodata & \nodata & \nodata \\
                       \nodata &  \nodata &             MS & $<$60 &             $<$13.56 &                                  \nodata & \nodata & \nodata & \nodata \\
       \alw\ $\lambda$1670.787 &        7 &           HVC3 & $<$90 &             $<$12.29 &                                  \nodata & \nodata & \nodata & \nodata \\
                       \nodata &  \nodata &            MS & $<$125 &             $<$12.43 &                                  \nodata & \nodata & \nodata & \nodata \\
       \few\ $\lambda$1144.939 &       50 &           HVC3 & $<$25 &             $<$13.30 &                                  \nodata & \nodata & \nodata & \nodata \\
                       \nodata &  \nodata &             MS & $<$25 &             $<$13.30 &                                  \nodata & \nodata & \nodata & \nodata \\
       \fet\ $\lambda$1122.526 &       51 &           HVC3 & $<$24 &             $<$13.60 &                                  \nodata & \nodata & \nodata & \nodata \\
                       \nodata &  \nodata &             MS & $<$33 &             $<$13.73 &                                  \nodata & \nodata & \nodata & \nodata \\
        \sw\ $\lambda$1259.519 &       18 &           HVC3 & $<$30 &             $<$14.13 &                                  \nodata & \nodata & \nodata & \nodata \\
                       \nodata &  \nodata &             MS & $<$33 &             $<$14.17 &                                  \nodata & \nodata & \nodata & \nodata \\
        \cf\ $\lambda$1548.205 &       10 &           HVC3 & $<$65 &             $<$13.21 &                                  \nodata & \nodata & \nodata & \nodata \\
                       \nodata &  \nodata &        MS & 120$\pm$18 &       13.57$\pm$0.07 &              $-$294$\pm$3 & 28$\pm$5 & 13.55$\pm$0.07 & 13.86$\pm$0.05 \\
                       \nodata &  \nodata &      HVC4 & 129$\pm$17 &       13.65$\pm$0.06 &                     $-$250$\pm$1 &  8$\pm$2 & 13.38$\pm$0.09 & \nodata \\
\nodata & \nodata & \nodata & \nodata & \nodata & $-$210$\pm$4 & 17$\pm$7 & 13.11$\pm$0.14 & \nodata \\
        \cf\ $\lambda$1550.778 &        6 &           HVC3 & $<$70 &             $<$13.54 &                                  \nodata & \nodata & \nodata & \nodata \\
                       \nodata &  \nodata &            MS & $<$112 &             $<$13.74 &                                  \nodata & \nodata & \nodata & \nodata \\
                       \nodata &  \nodata &          HVC4 & $<$107 &             $<$13.72 &                                  \nodata & \nodata & \nodata & \nodata \\
       \sif\ $\lambda$1393.755 &        9 &           HVC3 & $<$84 &             $<$12.97 &                                  \nodata & \nodata & \nodata & \nodata \\
                       \nodata &  \nodata &        MS &  49$\pm$16 &       12.80$\pm$0.12 &            $-$315$\pm$11 & 36$\pm$19 & 12.72$\pm$0.18 & 12.97$\pm$0.13 \\
                       \nodata &  \nodata &           HVC4 & $<$82 &             $<$12.96 &             $-$238$\pm$4 & 15$\pm$7 & [12.61$\pm$0.15]\tm{f} & \nodata \\
       \sif\ $\lambda$1402.770 &       10 &           HVC3 & $<$48 &             $<$13.02 &                                  \nodata & \nodata & \nodata & \nodata \\
                       \nodata &  \nodata &             MS & $<$51 &             $<$13.04 &                                  \nodata & \nodata & \nodata & \nodata \\
                       \nodata &  \nodata &           HVC4 & $<$39 &             $<$12.93 &                                  \nodata & \nodata & \nodata & \nodata \\
        \nf\ $\lambda$1238.821 &       13 &           HVC3 & $<$29 &             $<$13.13 &                                  \nodata & \nodata & \nodata & \nodata \\
                       \nodata &  \nodata &             MS & $<$28 &             $<$13.12 &                                  \nodata & \nodata & \nodata & \nodata \\
  \os\ $\lambda$1031.926\tm{j} &       46 &           HVC3 & $<$24 &             $<$13.29 &                                  \nodata & \nodata & \nodata & \nodata \\
                       \nodata &  \nodata &         MS &  74$\pm$7 &       13.84$\pm$0.02 &        $-$292$\pm$2 & 61$\pm$3\tm{i} & 14.03$\pm$0.02 & 14.03$\pm$0.02 \\
\enddata
\tn{a}{~Signal-to-noise ratio per resolution element in continuum next to line.\\}
\tn{b}{~AOD measurements were made separately in each HVC component, using the following LSR velocity ranges. HVC3: $-$450 to $-$360\kms; MS: $-$360 to $-$270\kms; HVC4: $-$270 to $-$190\kms. HVC4 is only seen in \sit, \sif, \cf, and \os\ (blended). For certain lines/components, the velocity ranges were slightly adjusted to avoid contamination or addition of noise, as follows: \few\ 1144, HVC3, $-$420 to $-$360\kms; \siw\ 1260, 1193 and \cw\ 1334, MS, $-$360 to $-$315\kms.\\}
\tn{c}{~Upper limits on $W_\lambda$ and log\,$N_a$ for non-detections (lines not detected at 3$\sigma$ significance) are calculated as $W_\lambda$(actual)+3$\sigma$ (e.g. if $W_\lambda$=20$\pm$10\,m\AA\ we quote $<$50\,m\AA). Lower limits are for saturated lines. Errors on  $W_\lambda$ and log\,$N_a$ include both statistical and continuum placement uncertainties (and 6\,m\AA\ fixed pattern noise for \fuse\ lines). When a component is not detected in multiple lines of a particular ion, only the strongest constraint on log\,$N$ is given.\\}
\tn{d}{~All lines of a given ion were fitted simultaneously, so VPFIT results are presented only once per ion. Two HVC3 components and two MS components were fitted for \sit. Two HVC4 components were fitted for \cf.\\}
\tn{e}{~Total VPFIT column density, summed over all fitted HVC components, with errors added in quadrature.\\}
\tn{f}{~VPFIT conducted even though absorption in this component not detected at 3$\sigma$ significance; results uncertain.\\}
\tn{g}{~No measurement of HVC absorption in \cw\ $\lambda$1036.337 made due to contamination with Galactic H$_2$ $\lambda$1035.183 and $\lambda$1035.427.\\}
\tn{h}{~No measurement of \ct\ $\lambda$977.020 made in HVC4 due to contamination with Galactic \oi\ $\lambda$976.448. Velocity centroids of the two MS \ct\ components tied to those of \sit\ during fit.\\}
\tn{i}{~Large $b$-value indicates unresolved sub-structure may be present.\\}
\tn{j}{~No measurement of \os\ $\lambda$1031.926 made in HVC4 due to contamination with Galactic H$_2$ $\lambda$1031.193.\\}
\end{deluxetable*}

\clearpage

\subsection{Voigt-Profile Fitting}
The {\sc VPFIT} software\footnotemark[7]\footnotetext[7]{Available at
  http://www.ast.cam.ac.uk/$\sim$rfc/vpfit.html.}  
was used to fit Voigt profiles to all lines showing MS detections,
yielding the column densities, line widths, and velocity
centroids of absorption. The code acts simultaneously on multiple
lines from the same species (e.g. doublets), and accounts for
instrumental broadening by convolving the model spectra with the
instrumental line spread function (LSF) during the fit.
The LSF is assumed to be Gaussian, with FWHM=20\kms\ for \fuse,
7.0\kms\ for STIS/E140M, and 7.5\kms\ for VLT/UVES. 
The number of components to fit has to be specified manually.
With one exception (for \siw\ and \sit\ toward \qa; see \S4.1)
we did not tie components together between different species, since
{\it a priori} it is not known which species 
co-reside in the same volumes. The total VPFIT column density
presented for each ion is a straight summation across all HVC components. 
The VPFIT errors presented in Tables 1 and 2 represent the statistical
uncertainties only, and not continuum placement uncertainties or the
systematic errors associated with fitting simple Voigt profiles to
complex velocity structures.

\subsection{Apparent Optical Depth (AOD) Measurements}
The AOD method \citep{SS91, SS92, Je96} is a technique used to derive
column densities within a specified velocity range $v_-$ to
$v_+$. The results are accurate provided the lines are resolved and
unsaturated, and that the data have S/N$\ga$8 \citep{Fo05b}.
Unresolved saturation can be difficult to detect
(affecting both VPFIT and AOD measurements). 
The velocity ranges were determined by visual inspection of the data,
finding the velocities where the flux recovers to the
continuum on either side of the HVC absorption components.
The AOD in each pixel is defined by $\tau_a(v)$=ln\,$[F_c(v)/F(v)]$,
where $F(v)$ and $F_c(v)$ are the observed flux and the estimated
continuum flux, respectively. The total AOD is given by 
$\tau_a=\int_{v_-}^{v_+}\tau_a(v){\mathrm d}v$, and
the apparent column density
$N_{\rm a}(v)=3.768\times10^{14}(f\lambda)^{-1}\tau_a(v)$.
Both continuum errors and statistical errors are included in the error
analysis, and for lines measured in \fuse\ data, a fixed pattern noise of
6\,m\AA\ was added in quadrature. 
We adopt the AOD column densities for use in the rest of the paper, 
since they are less sensitive to the choice of components and
have a more thorough error analysis, 
though a comparison of the VPFIT and AOD column densities 
shows generally consistent results (see Tables 1 and 2).

For non-detections, i.e. cases where there is no absorption
present at 3$\sigma$ significance in the chosen velocity range, we
determined the 3$\sigma$ equivalent width limit; for example, if a
measurement found $W_{\lambda}$=20$\pm$10\,m\AA, we 
adopted $W_{\lambda}^{3\sigma}\!<\!50$\,m\AA\ (since this is the largest
value that $W_{\lambda}$ can take, at 99.7\% significance). 
We then converted $W_{\lambda}^{3\sigma}$ into a 3$\sigma$ limit
on the column density assuming a linear curve-of-growth.
For saturated lines, which in practice we define as those where 
the normalized flux falls below 0.05 at any point in the line profile, 
we take the returned AOD value of $N_{\rm a}$ but treat it as a
lower limit. 

\section{Overview of Magellanic Stream Absorption}
Figures 2 and 3 show the normalized UV absorption-line profiles toward
\qa\ and \qb, respectively. Included are all absorption lines showing
MS detections, and several other lines showing MS non-detections that
can be used to derive upper limits on the ionic column density.
We also include 21\,cm emission spectra and, in the case
of \qa, optical spectra of \caw\ and \nao.
A HV component is observed near \vlsr=$-$340\kms\ in both sightlines,
matching the velocity of the \hi\ 21\,cm emission from the MS toward \qa.
We thus identify the $-$340\kms\ component as the MS.
Also visible in both sight-lines are other HVCs and the Galactic absorption
components at $|v_{\rm LSR}|\lesssim$100\kms.

\begin{figure*}
\epsscale{1.1} 
\plotone{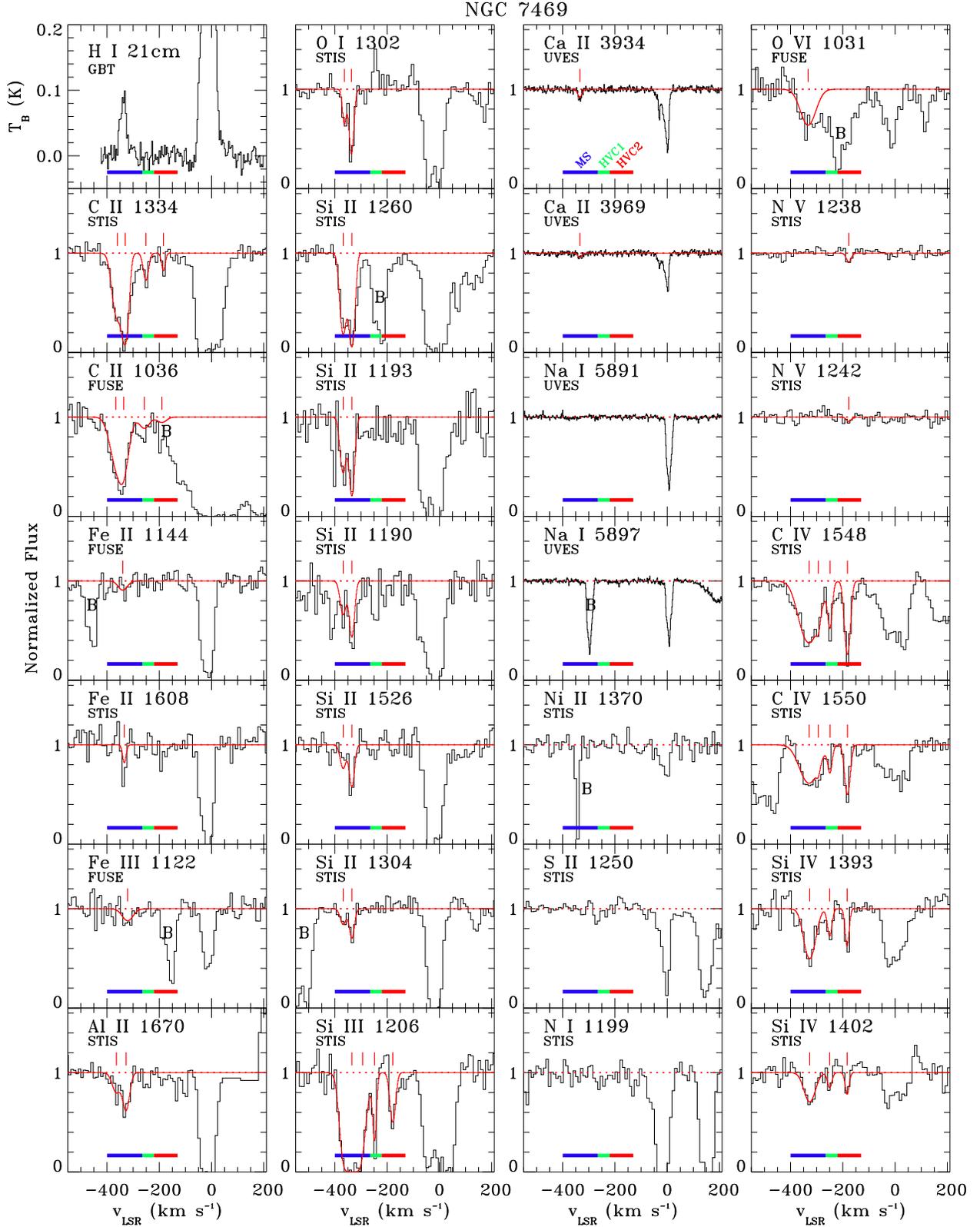}
\caption{STIS, \fuse, and UVES absorption-line profiles 
  and GBT 21\,cm emission-line profile toward NGC~7469. 
  Normalized flux is plotted against LSR velocity for each absorption 
  line shown. The 21\,cm panel shows the brightness temperature
  profile. The \fuse, STIS, UVES, and GBT data have been rebinned by
  five, three, two, and five pixels, respectively. 
  MS absorption is identified 
  at \vlsr=$-$340\kms. Two other HVCs are seen at
  $-$260\kms\ (HVC1) and $-$190\kms\ (HVC2) in some species.
  The colored bars in each panel show the velocity ranges used in the
  AOD integrations (MS in blue).
  Voigt profile fits (on a 1\kms\ grid) to the HVC absorption 
  are shown in red, with tick marks identifying component centers.
  The label 'B' denotes a blend.} 
\end{figure*}

\begin{figure*}
\epsscale{1.1} 
\plotone{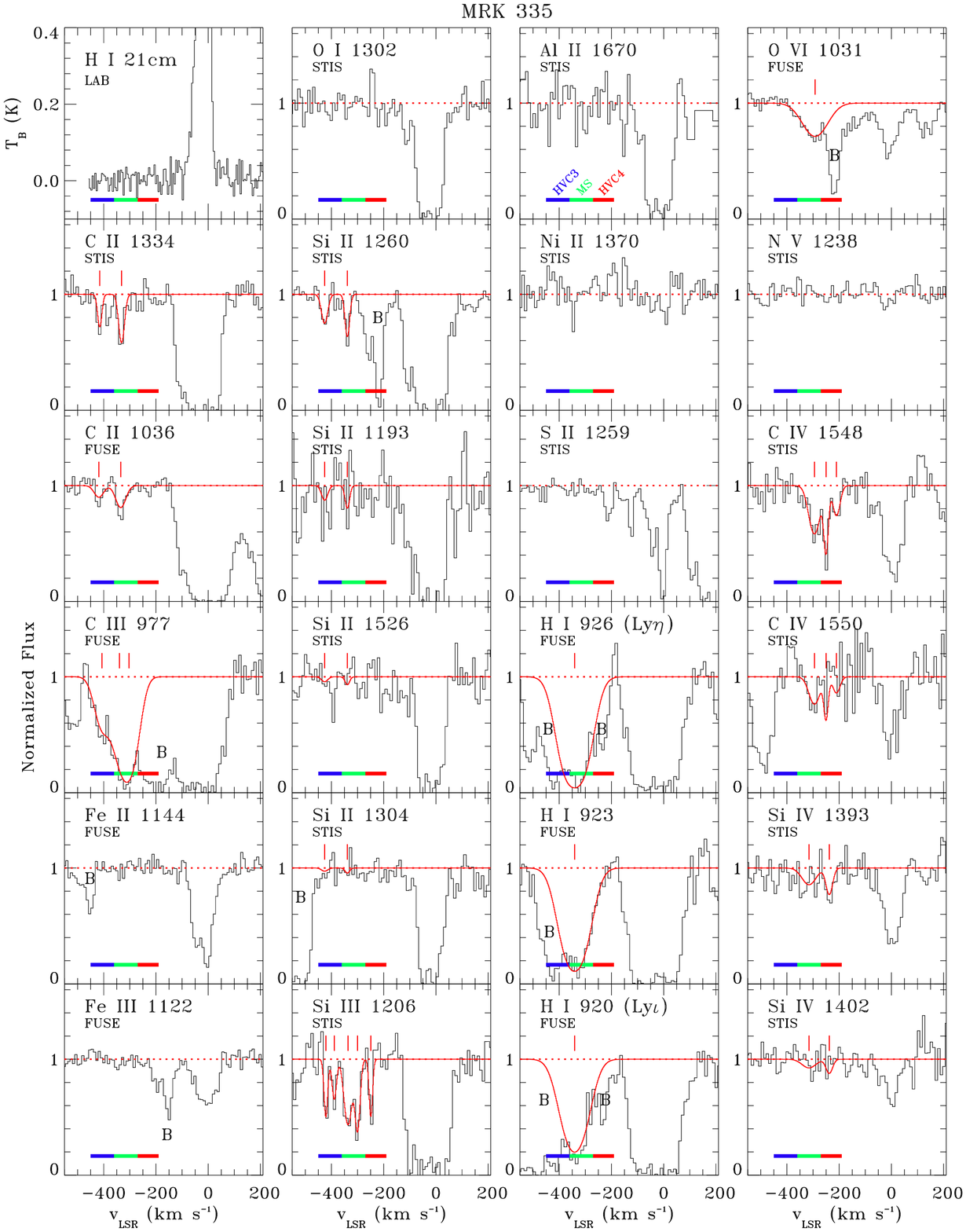}
\caption{STIS and \fuse\ absorption-line profiles,
  and LAB 21\,cm emission-line profile toward Mrk~335. Normalized flux
  is plotted against LSR velocity for each absorption line shown.
  The 21\,cm panel shows the brightness temperature profile. 
  The \fuse, STIS, and LAB data have been rebinned by five, three, and
  five pixels, respectively. 
  No 21\,cm emission from the MS is seen, but we identify the
  \vlsr=$-$340\kms\ absorption component as the MS.
  Two other HVCs are seen at $-$420\kms\ (HVC3) and
  $-$250\kms\ (HVC4) in some ions.
  The colored bars in each panel show the velocity ranges used in the
  AOD integrations (MS in green).
  Voigt profile fits (on a 1\kms\ grid) to the HVC absorption 
  are shown in red, with tick marks identifying component centers.
  The label 'B' denotes a blend.}
\end{figure*}

\subsection{MS Absorption toward NGC~7469}
Toward \qa, the 21\,cm observations from the GBT, Green Bank
140\,foot, and Effelsberg telescope all show the MS in emission near
$-$335\kms. The MS component is not detected in the LAB survey,
but this can be attributed to that survey's larger beam size
(35\arcmin). Measurements of the \hi\ emission in these four datasets are
given in Table 3. The dispersion between the derived values of
$N$(\hi) is a consequence of the differing beam sizes,
and reflects the small-scale structure known to exist in HVCs on
scales below 10\arcmin\ \citep{Sa00, Wa01, Se04, Br10}. 
For our final value of log\,$N$(\hi),
we adopt a weighted mean of the GBT and Effelsberg values
(the two observations with the smallest beam size, 
9.1\arcmin\ and 9.7\arcmin, respectively).
We also include a systematic error equal to the 1$\sigma$ dispersion
between the three measurements of log\,$N$(\hi) listed in Table 3, to account
for small-scale structure. This gives our adopted value 
log\,$N$(\hi)=18.63$\pm$0.03(stat)$\pm$0.08(syst). 
\begin{deluxetable*}{lcccc ccc}
\tablewidth{0pt}
\tabletypesize{\small}
\tablecaption{21\,cm Observations of the MS Toward NGC~7469}
\tablehead{Radio Telescope & Beam Size & Sensitivity & Resolution &
  log\,$N$(\hi) & $v_0$(\hi) & $b$(\hi) & Reference\\
& (arcmin) & (mK) & (\kms) & ($N$ in cm$^{-2}$) & (\kms) & (\kms) & }
\startdata
Green Bank Telescope (GBT) & 9.1 & 20 & 1.0 & 18.67$\pm$0.03 & $-$337
& 16.3 & Brown et al. 2010\\
Effelsberg 100\,m          & 9.7 & 20 & 1.0 & 18.54$\pm$0.05 & $-$332
& 16.0 & Wakker et al. 2001\\
Green Bank 140\,ft         & 21  & 10 & 2.0 & 18.52$\pm$0.06 & $-$342
& 26.6 & Brown et al. 2010\\
Leiden-Argentine-Bonn (LAB)& 35  & 60 & 1.0 & $<$18.52       &\nodata
&\nodata & Kalberla et al. 2005 
\enddata
\tablecomments{We adopt a weighted mean of the GBT and Effelsberg
  column densities, because of their similar small beam size. 
  We also include a systematic error equal to the
  1$\sigma$ dispersion between the three values of log\,$N$(\hi)
  listed above to account for 
  small-scale structure. Our final adopted column density   
  is log\,$N$(\hi)=18.63$\pm$0.03(stat)$\pm$0.08(syst).}
\end{deluxetable*}
\normalsize

HV absorption is identified in the spectra of \qa\
in the velocity range $-$400 to $-$130\kms\ in the low-ion lines
\oi\ $\lambda$1302, \cw\ $\lambda\lambda$1036,1334, 
\siw\ $\lambda$$\lambda$1260,1193,1190,1526,1304, \sit\ $\lambda$1206, 
\alw\ $\lambda$1670, and \caw\ $\lambda\lambda$3934,3969.
Low significance absorption is seen in \few\
$\lambda$1144 (1.2$\sigma$ detection) and $\lambda$1608 (2.1$\sigma$), 
and in \fet\ $\lambda$1122 (2.5$\sigma$). 
Among the high ions, \os\ $\lambda$1031, 
\cf\ $\lambda\lambda$1548,1550, and \sif\ $\lambda\lambda$1393,1402
are clearly detected, whereas absorption in
\nf\ $\lambda$1238 and $\lambda$1242 is detected at
2.4$\sigma$ and 1.8$\sigma$ significance, respectively, but only in a
single component centered near $-$177\kms, not in the main MS
component at $-$340\kms. 
No significant HV absorption is detected in
\nao\ $\lambda\lambda$5891,5897, \sw\ $\lambda\lambda$1259,1253,1250 
or \no\ $\lambda\lambda$1199,1200.2,1200.7.
The lines \os\ $\lambda$1037, \oi\ $\lambda$1039 and \cw$^*$
$\lambda$1335 are blended in the velocity interval $-$400 to $-$200\kms,
so cannot be measured. 
\ct\ $\lambda$977 and the higher-order \hi\ Lyman series lines, lying
in the \fuse\ SiC channels below 1\,000\,\AA, are too noisy for any
useful measurement. 

Analysis of the profiles with the highest S/N ratio
(\cw, \sit\ and \cf) reveals at least three HV components. These are
identified with colored horizontal bars on Figure 2 and defined as follows:
the main MS component covering the range $-$400 to $-$265\kms\
and centered at $-$340\kms, 
a weaker component (which we label HVC1) 
covering $-$265 to $-$220\kms\ and centered at $-$250\kms, and 
a third component (HVC2) 
covering $-$220 to $-$130\kms\ and centered at $-$200\kms. 
In \oi, \cw, \siw\ $\lambda$1260 and \alw,
the main MS component shows two sub-components at $-$365 and
$-$335\kms. For \sit, complete saturation in the range 
$-$360 to $-$305\kms\ prevents us from resolving these
sub-components, so we do not attempt a fit to the MS components in
\sit, though we note that in the unsaturated interval $-$400 to
$-$360\kms, the \sit\ profile is consistent with that of \siw,
suggesting the two ions are co-spatial at these velocities.

The HVC component structure can be seen in more detail
in the left panels of Figure 4, where we plot the apparent column
density profiles versus velocity for the low ions \cw, \caw, \siw, and
\sit, and the high ions \cf, \nf, and \os.
With the exception of \nf, the high-ion profiles show a broadly
similar component structure as the low-ion profiles, but with
different relative strengths. 
Furthermore, the \cf\ profile shows an extra component near
$-$295\kms, not seen in the low ions (expect possibly in \sit).
In the bottom-left panel of Figure 4,
we show the \cf/\cw\ column density ratio versus velocity,
which indicates the relative amounts of high-ion and low-ion gas. 
This ratio was formed by interpolating the \cf\
profile onto the same velocity grid as the \cw\ profile, and
propagating the errors on $N_{\rm a}$(\cf) and $N_{\rm a}$(\cw).
HVC1 and HVC2 exhibit markedly higher \cf/\cw\ ratios 
than the principal MS component: 
log $N_{\rm a}$(\cf)/$N_{\rm a}$(\cw) changes
by over 1\,dex from $\approx\!-0.7$ in the MS to $\approx\!+0.5$ in HVC2, 
indicating that relatively more high-ion gas is present in HVC1 and
HVC2 than in the MS component. Note that \citet{Se03} classified HVC1
and HVC2 as ``LG'' (Local Group) absorbers, which could account for
their different ionization level. It is also possible that they are
ionized fragments of the outer Stream that have acquired velocity
offsets relative to the MS, and are in the process of
being evaporated through the MS/corona interaction.

\begin{figure*}
\epsscale{1.0}\plotone{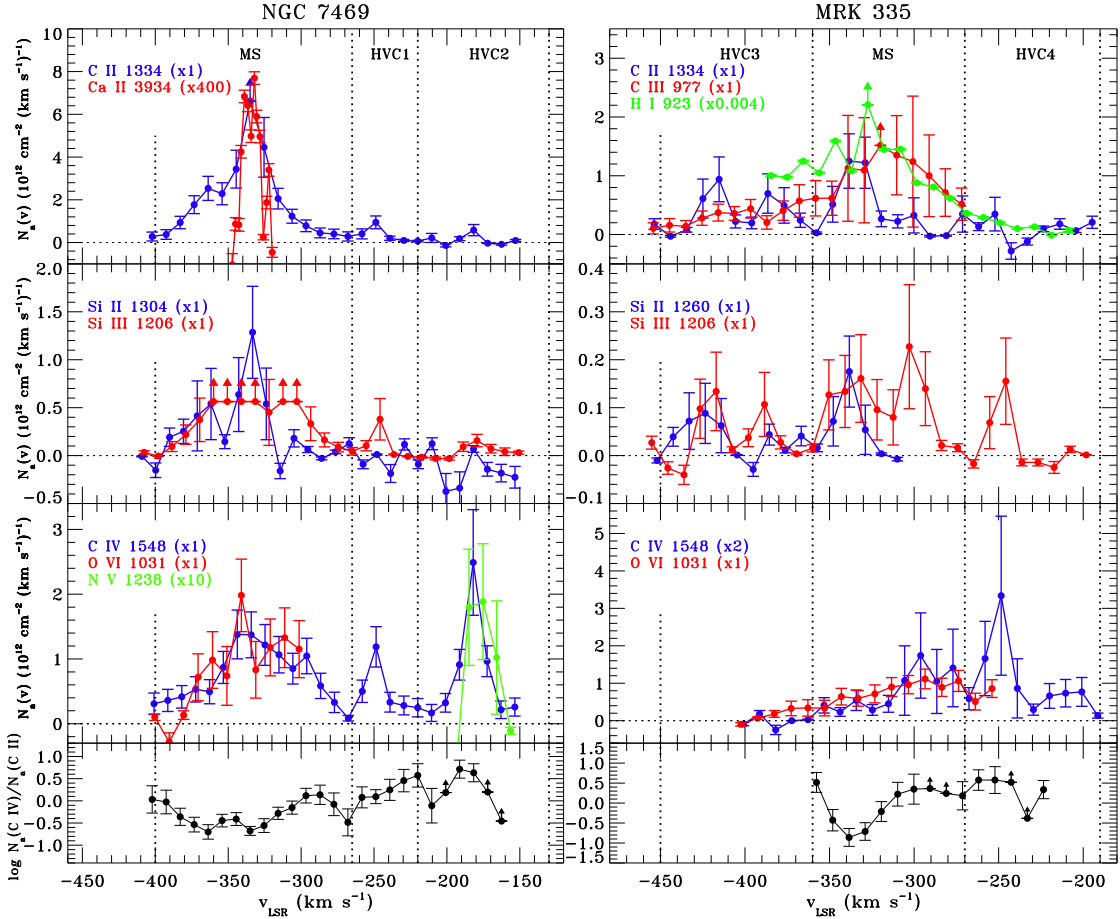}
\caption{Apparent column density profiles of high-velocity
  absorption toward NGC~7469 (left) and Mrk~335 (right). 
  The profiles of some ions have been scaled by the factors annotated
  in the legend to ease inter-comparison.
  The velocity integration ranges for each component are shown with
  dotted vertical lines.
  The profiles are only shown over unblended velocity ranges, and
  saturated pixels are shown with lower-limit arrows.
  The bottom panels show
  the \cf/\cw\ apparent column density ratio vs velocity, indicating
  the relative amounts of high-ion and low-ion gas.}
\end{figure*}

\subsection{MS Absorption toward Mrk~335}
Toward \qb, the best 21\,cm data available covering MS velocities
are from the LAB survey. These data show no detection of the Stream
(Figure 3); we measure a 3$\sigma$ limit
log\,$N$(\hi)$_{\rm MS}<$18.35 in the range $-$400 to $-$300\kms.
However, we determined $N$(\hi)$_{\rm MS}$ 
toward \qb\ by fitting the $-$340\kms\ component seen in the
higher-order Lyman lines in the \fuse\ data. 
Using a simultaneous single-component VPFIT to \hi\ $\lambda$926 (Ly$\eta$),
$\lambda$923 (Ly$\theta$), and $\lambda$920 (Ly$\iota$), we derive
log\,$N$(\hi)=16.67$\pm$0.05. 
We tried fitting these three \hi\ lines with a two-component MS model
(as seen in \sit; see below), but
VPFIT rejected the second component, i.e. a more complex \hi\
component structure is not warranted by the data quality, and we
proceed with a single-component fit.
For reasons of space, only three \hi\ lines are shown in Figure
3; the full Lyman series is displayed in Figure 2 of \citet{Fo06}. 

The \qb\ STIS spectrum is noisier than the \qa\ STIS spectrum, but
still shows HV absorption in the LSR velocity range  
$-$450 to $-$200\kms\ in \cw\ $\lambda\lambda$1036,1334,
\siw\ $\lambda$$\lambda$1260, \sit\ $\lambda$1206,
\cf\ $\lambda\lambda$1548,1550, and \sif\ $\lambda\lambda$1393,1402. 
The \fuse\ spectrum shows clear HV absorption in \ct\ $\lambda$977
and \os\ $\lambda$1031. No significant HV absorption is detected in
\oi\ $\lambda$1302, \siw\ $\lambda$$\lambda$1193,1526,1304,
\alw\ $\lambda$1670, \niw\ $\lambda$1370,
\sw\ $\lambda\lambda$1259,1253,1250,
\no\ $\lambda\lambda$1199,1200.2,1200.7, or \nf\ $\lambda\lambda$1238,1242.
We do not possess an optical spectrum of this object, 
so no \caw\ or \nao\ profiles are available.

Two narrow HV components are observed in the low ions \cw\ and \siw,
the first centered at $-$415\kms\ and the second at $-$330\kms.
Based on the agreement with the 21\,cm velocity
of the MS toward \qa\ (24.9\dgr\ away), we identify the $-$330\kms\
component as the MS, and label the $-$415\kms\ feature as HVC3,
although we cannot rule out the possibility that both components are
related to the MS.
For our AOD integrations we define HVC3 
from $-$450 to $-$365\kms, and the MS from $-$365 to $-$315\kms. 
Both components are also seen in \ct\ and \sit, although
in these two ions the MS absorption extends further to $-$270\kms.
In \sit\ both the HVC3 and MS intervals show two sub-components,
which we fit individually, though higher S/N 
observations are needed to confirm this sub-structure.
We fit the \ct\ MS absorption with a two-component model with
velocity centroids tied to the values obtained from the \sit\ fit.
 
The high ions show a different component structure than the
low ions. No high-ion absorption is detected in HVC3, but
\cf, \sif, and \os\ all show clear absorption in the MS velocity range.
The high-ion absorption peaks near $-$300\kms, as opposed to
$-$340\kms\ for the low ions (note the $-$300\kms\ high-ion component
is also seen toward \qa).
Furthermore, \cf\ and \sif\ show additional HV absorption in
the range $-$270 to $-$190\kms, which we label HVC4 (\os\ is present
but blended with H$_2$ at these velocities). None of the low ions 
(except \sit) show detectable absorption in HVC4.
The \cf\ $\lambda$1548 profile shows two components within the HVC4 range. 
The differing behaviour of the high- and
low-ions toward \qb\ can be seen in the 
apparent column density profiles shown in the right panels of Figure 4.
As in the case of \qa, the logarithmic \cf/\cw\ ratio changes from 
$\approx\!-0.7$ at MS velocities to $\approx\!0.5$ in HVC4.

\section{Metallicity of the MS}
Toward \qa, we can directly determine the oxygen abundance in the MS
using the measured column densities of \oi\ and \hi.
Because of the similar ionization potentials of O and H, 
a charge exchange reaction couples the two species together
\citep{FS71, Vi95}. Furthermore oxygen is typically not strongly 
depleted onto interstellar dust \citep{Me98, Je05}. 
Thus the observed ratio [\oi/\hi] gives a close
indication of the true oxygen abundance [O/H] and the overall metallicity. 
Combining the measured 
log\,$N$(\hi)=18.63$\pm$0.03(stat)$\pm$0.08(syst) from the 21\,cm data with 
log\,$N$(\oi)=14.32$\pm$0.04 from \oi\ $\lambda$1302 
and the solar oxygen abundance log\,(O/H)$_\odot$=$-$3.31, we find
[O/H]$_{\rm MS}$=$-$1.00$\pm$0.05(stat)$\pm$0.08(syst), 
where we use the solar (photospheric) abundances of \citet{As09} and the
standard notation [X/Y]=log\,($N_{\rm X}$/$N_{\rm Y}$)--log\,(X/Y)$_\odot$.
This is 
closer to the SMC interstellar oxygen abundance,
[O/H]$_{\rm SMC}$=$-$0.66$\pm$0.10, than to the LMC value, 
[O/H]$_{\rm LMC}$=$-$0.34$\pm$0.06
\citep[][updated to the latest solar abundances]{RD92},
and close to the oxygen abundance in the Magellanic Bridge
[O/H]$_{\rm MB}$=$-$0.96$^{+0.13}_{-0.11}$ measured by \citet{Le08}.
Our value for [O/H]$_{\rm MS}$ is lower than previous MS
metallicity estimates \citep{Lu98, Gi00, Se01} by a factor 
of $\approx$2--4, but we emphasize
that our measurement is the first derived from the
\oi/\hi\ ratio, which is robust against ionization corrections, as
opposed to the \siw/\hi\ or \sw/\hi\ ratio, which are not.
Our value supports the view that the Stream originates in the SMC
rather than the (higher-metallicity) LMC.
Using the non-detection of \no\ $\lambda$1199.5496 in the MS toward
\qa, we measure [N/H]$_{\rm MS}$=[\no/\hi]$<$--0.92 at 1$\sigma$ and 
$<$--0.44 at 3$\sigma$ confidence, 
consistent with the interstellar nitrogen abundance in the SMC,
[N/H]$_{\rm SMC}$=$-$1.20$\pm$0.20 \citep{RD92},
though a higher S/N spectrum is needed to place a stronger limit.

\section{Low Ions: Photoionization Modeling}
In this section we investigate the physical conditions and chemical
abundances in the MS by modeling the metal-line column densities
observed toward \qa\ and \qb\ with the photoionization code CLOUDY.

\subsection{CLOUDY Methodology}
We ran a photoionization model to the observed low-ion column
densities in the MS using the one-dimensional photoionization code
CLOUDY \citep[v08.00; last described in][]{Fe98}. 
The goal of this process is to solve for the ionization parameter
log\,$U$ in the low-ion phase, where $U\!\equiv\!n_\gamma/n_{\mathrm H}$,
the ratio of ionizing photon density to gas density,
which in turn determines the density and line-of-sight size of the cloud.
Our CLOUDY runs assume the HVCs are plane-parallel uniform-density
slabs exposed to an ionizing radiation field. We compute this field as
the sum of two contributions:\\
(a) the $z$=0 extragalactic background (EGB) incorporated within
CLOUDY. The EGB is based on \citet{HM96, HM01} but updated to include
the contribution from starburst galaxies, and has a specific flux at 1 Ryd of
$F_{912}$=$10^{-21.32}$\,erg\,cm$^{-2}$\,s$^{-1}$\,Hz$^{-1}$ 
\citep[see also][]{Sh99}, and a total ionizing flux  
$\Phi_{\rm EGB}$=$10^{4.5}$\,photons\,cm$^{-2}$\,s$^{-1}$, and\\ 
(b) the radiation field escaping from the Milky Way
(MW), as parameterized in \citet{Fo05a} based on the model of
\citet{BM99}, with an escape fraction of 6\% normal to the disk. 
This Galactic contribution is calculated assuming an
MS distance of 50\,kpc, giving a specific flux at 1 Ryd of
$F_{912}$=$10^{-21.33}$\,erg\,cm$^{-2}$\,s$^{-1}$\,Hz$^{-1}$, 
and a total ionizing photon flux 
$\Phi_{\rm MW}$=$10^{4.5}$\,photons\,cm$^{-2}$\,s$^{-1}$.\\ 
The total ionizing fluxes in the MW (at 50 kpc) and EGB fields
are very similar, but the shallower (i.e. harder) slope of the EGB
ensures it dominates the provision of high-energy photons.
The ionizing photon density $n_\gamma$ in our combined MW+EGB field
is $10^{-5.85}$\,cm$^{-3}$. 

The following procedure was used to solve for log\,$U$ and 
the relative abundance pattern in the MS along each sight line.
We ran a grid of CLOUDY models at different ionization parameter, with
log\,$U$ varied from $-$1.0 to $-$5.0 in 0.05\,dex intervals, taking
the \hi\ column density and the metallicity [O/H]$_{\rm MS}$=$-$1.00
as fixed inputs. We then found the model (i.e. the value of log\,$U$)
that best reproduced the observations, by
minimizing the squared residuals between the observed and predicted
column densities for the ions included in the model. 
The included ions are only those
for which successive ion stages from the same element are available
(\siw/\sit, \cw/\ct, or \few/\fet).
These successive ion stages are the key to the CLOUDY models,
since for a given $N$(\hi), a given ratio uniquely specifies log\,$U$.
We then find the non-solar \emph{relative} abundances
(if any) that are needed to reproduce the \emph{absolute} column
densities of all low ions detected in the MS.
The entire process is then repeated for each sight line.

The validity of the CLOUDY models relies crucially on the assumption
that the ions included are co-spatial. We assess this in each sight
line by detailed comparison of the velocity component structure. 
In addition, the \qa\ model assumes the \hi\ column density along the
pencil-beam line-of-sight is equal to the value derived from a 21\,cm
observation taken with a 9.1\arcmin\ beam (see discussion in \S4.1).
The \qb\ model does not suffer from this uncertainty since in this
sightline $N$(\hi) is derived from the Lyman series absorption lines.
Finally, uncertainty in the distance to the MS 
translates into uncertainty in the flux of Galactic ionizing photons
incident on the clouds. As an approximate guide, a 10\% (5\,kpc)
error in the MS distance translates to a $\sim$20\% error
in the Galactic ionizing flux. 
Such an error would not change the derived
value of log\,$U$, but if the MS is closer (further) than 50\,kpc, the
derived gas density will be higher (lower) than the
values reported below, since $n_\gamma$ will be higher (lower).

\begin{figure*}
\epsscale{1.1}\plottwo{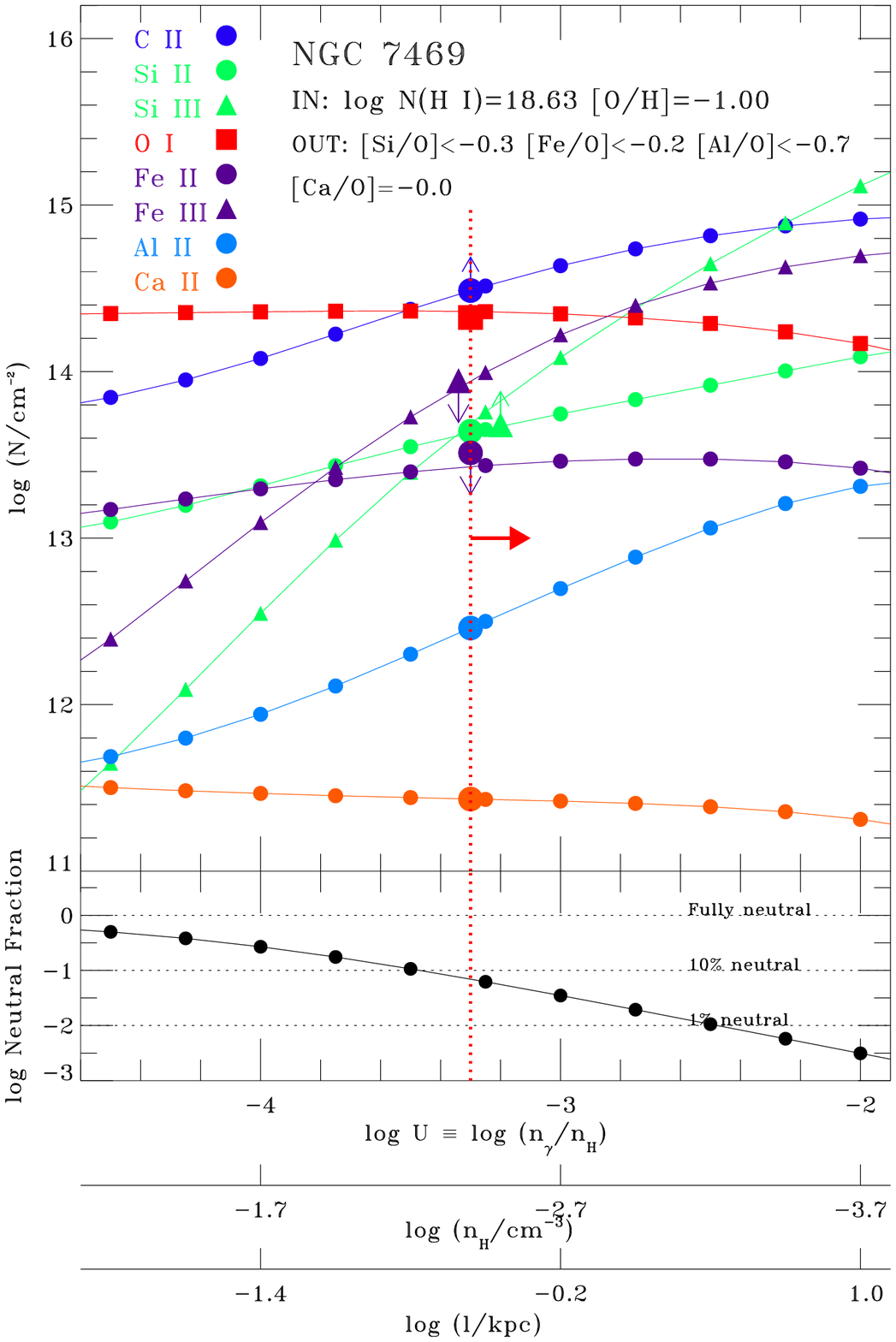}{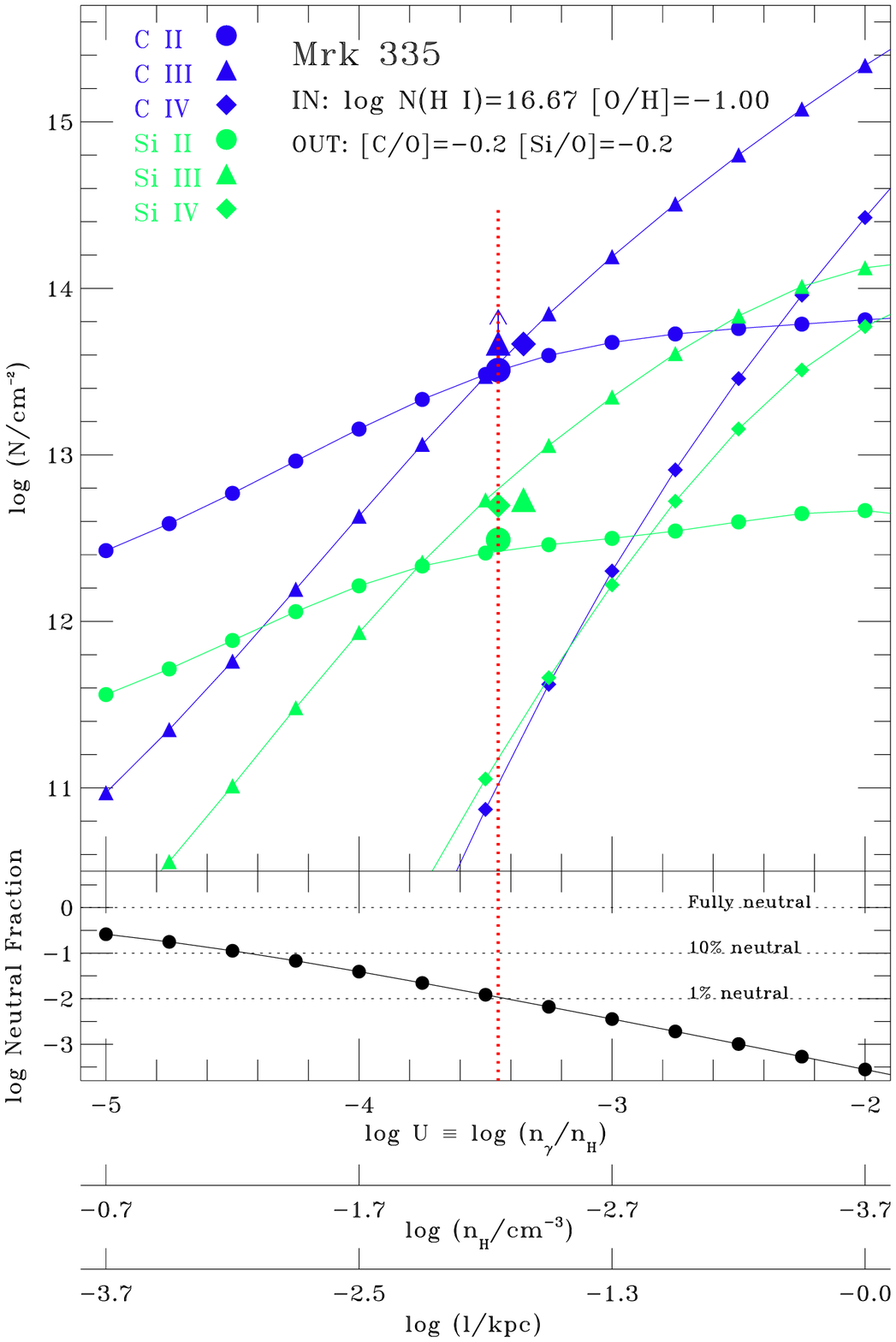}
\caption{CLOUDY photoionization models of the low-ion phase of the
  MS toward NGC~7469 (left) and Mrk~335 (right). 
  In the upper panels, the predicted column density is plotted against
  ionization parameter $U$ for several detected ions
  (color-coded in the legend).
  Each curve has been scaled according to the relative abundance
  pattern found by the model.
  The large data points show the observations, plotted at the
  best-fit log\,$U$ (shown with a dotted line);
  the \sit\ and \cf\ points have a small offset introduced in the
  x-direction for clarity. 
  The lower panels shows the logarithm of the neutral hydrogen fraction 
  $x_{\rm H I}$=$N$(\hi)/$N$(\hi+\hw) in the model, and the bottom two
  axes show the gas density and line-of-sight cloud size in kpc.
  The NGC~7469 model gives log\,$U\!>\!-$3.30 (only a limit since
  \sit\ is saturated), which in turn constrains [Si/O], [Fe/O],
  [Al/O], and [Ca/O].
  The best-fit Mrk~335 model has log\,$U$=$-$3.45$\pm$0.08, corresponding
  to a hydrogen ionization level of 98.9$\pm$0.2\%; this model does
  not explain the \cf\ or \sif\ column densities, which must arise in
  a separate phase.}
\end{figure*}

\subsection{CLOUDY Results: NGC~7469 Sight Line}
Toward \qa, the lines of successive ionization states covered by the
data are \few/\fet\ and \siw/\sit. 
\few\ and \fet\ are each only marginally detected and their ratio is
not reliably determined, so they were not used to constrain log\,$U$.
A good \siw\ MS column density is available; we adopt the value 
log\,$N_a$(\siw)=13.64$\pm$0.05 measured from the 1193 line, but due to
saturation in \sit\ between $-$360 and $-$300\kms\ we can only derive
a lower limit on the MS \sit\ column, log\,$N_a$(\sit)$>$13.82.
Applying our CLOUDY model to the limit on the MS \sit/\siw\ ratio, 
log\,[$N$(\sit)/$N$(\siw)]$>$0.26, yields log\,$U\!>\!-$3.15. 
However, looking closely at the line profiles, we see that \sit\ has a
contribution from the $-$300\kms\ component seen in \cf\ and \os\ but
not in \siw
This extra absorption should not be included in the CLOUDY model of
the low-ion phase, so we repeated the \sit\ AOD measurement in the
range $-$400 to $-$315\kms, where the \siw\ and \sit\ profiles are 
consistent. This gives log\,$N_a$(\sit)$>$13.68 and
log\,[$N$(\sit)/$N$(\siw)]$>$0.12, in turn giving log\,$U\!>\!-$3.30. 
This limit on log\,$U$ translates to a gas density 
log\,($n_{\rm H}$/cm$^{-3}$)$<-$2.36, 
a neutral hydrogen fraction 
$x_{\rm H}<0.07$ ($>$93\% hydrogen ionization level), 
a total hydrogen column density log\,$N$(\hi+\hw)$>$19.79, 
and a line-of-sight cloud size $l$=$N$(\hi+\hw)/$n_{\rm H}\!>$4.6\,kpc.
This model is shown on the left panel of Figure 5.

The lower limit log\,$U\!>\!-3.30$ allows us to constrain the abundances of
Si, Fe, Al, Ca. We find [Si/O]$_{\rm MS}\!<\!-0.3$, 
[Fe/O]$_{\rm MS}\!<\!-0.2$,  [Al/O]$_{\rm MS}\!<\!-0.7$, 
and [Ca/O]$_{\rm MS}$=0.0$\pm$0.1, i.e.
we derive sub-solar abundances of Si, Fe, and Al (relative to O), but 
an approximately solar Ca relative abundance.
The results for Si/O, Fe/O, and Al/O are all upper limits since the
solution to log\,$U$ must be above $-$3.30, and in this regime,
the curves showing the predicted \siw, \few, and \alw\ columns 
on Figure 5a must move \emph{down} to reproduce the 
corresponding observations. 
The result for Ca/O is a value (not a limit)
because \caw\ and \oi\ track each other closely at all values of
log\,$U$ under consideration, both with almost flat curves. 
Since Si and O are both $\alpha$-elements,
they share a common nucleosynthetic origin and their intrinsic ratio
should be close to solar, so \emph{the sub-solar value of 
[Si/O]$_{\rm MS}$ indicates that Si is depleted into dust grains} 
(O being undepleted), by an amount 
at least as large as the typical Si depletion 
in Milky Way halo clouds \citep[$-$0.26$\pm$0.14;][]{SS96}.
The depletions of Al and Fe are consistent with the
presence of dust in the MS.
The finding that there is little room for depletion of Ca onto dust grains
is puzzling, but mirrors a similar result found by \citet{Ri09} in
their study of \oi\ HVCs (in non-MS directions).

The presence of dust in the MS has been inferred before, by
(a) the detection of molecular hydrogen, which forms on the surface of
dust grains, toward NGC~3783 \citep{Se01} and Fairall~9 \citep{Ri01}, and
(b) the super-solar sulfur-to-iron ratio of 
S/Fe=7.6$\pm$2.2 solar measured in the MS toward NGC~3783 by \citet{Lu98},
indicating that Fe (but not S) is depleted out of the gas phase into dust.
However, in the \qa\ sightline under study here, 
\citet{Wa06} found \emph{no} H$_2$ in the MS 
(only a low-velocity, Galactic component), indicating that
the neutral gas column [log\,$N$(\hi)$_{\rm MS}$=18.63] is
too low to create a detectable H$_2$ column.

\subsection{CLOUDY Results: Mrk~335 Sight Line}
Toward \qb, the available single-element diagnostic ratios are
\sit/\siw\ and \ct/\cw\ (\few\ and \fet\ are undetected).
Because of saturation in \ct\ $\lambda$977, the measured \ct/\cw\ ratio
is a lower limit, but neither \siw\ nor \sit\ appears
saturated, so a solution for log\,$U$ is possible from the
\sit/\siw\ ratio. The key question is whether 
the \hi\ and the singly- and doubly-ionized species are co-spatial. 
Looking at the line profiles,
the \ct\ and \hi\ column density profiles track each
other fairly well over MS velocities (Figure 4, top-right panel).
However, the MS component appears broader in \ct\ 
($b$=37$\pm$12\kms) and \hi\ ($b$=58$\pm$2\kms) 
than in \cw\ ($b$=12$\pm$2\kms) and \siw\ ($b$=9$\pm$2\kms), 
and the \sit\ profile (observed at 7\kms\ resolution)
shows sub-structure in the MS velocity interval.
Given the lower \fuse\ resolution below 1\,000\AA\ and
the presence of saturation, this indicates that unresolved component
structure is likely present in \ct\ and \hi. 

If we assume that \cw, \ct, \siw, \sit, and \hi\ co-reside in the same gas, 
then using the column densities measured in the range 
$-$360 to $-$270\kms, where log\,[$N$(\sit)/$N$(\siw)]=0.52$\pm$0.08 
and log\,[$N$(\ct)/$N$(\cw)]$>$0.49, we derive 
log\,$U$=$-$3.15$\pm$0.08, [C/O]=$-$0.3$\pm$0.1 and
[Si/O]=$-$0.3$\pm$0.1.
However, this solution does not account for the contribution to the
\sit\ column from the highly-ionized $-$300\kms\ component seen in
\cf\ and \os\ but not in \siw. 
To account for this, we repeated our \sit/\siw\ and \ct/\cw\
measurements in the restricted interval $-$360 to $-$315\kms, where
the \siw\ and \sit\ profiles closely overlap and where there is little
high-ion absorption.
In this interval we find log\,[$N$(\sit)/$N$(\siw)]=0.24$\pm$0.09 and 
log\,[$N$(\ct)/$N$(\cw)]$>$0.16. 
Using these revised ratios in the CLOUDY model yields
log\,$U$=$-$3.45$\pm$0.08, [C/O]=$-$0.2$\pm$0.1 and [Si/O]=$-$0.2$\pm$0.1. 
This model is shown in the right panel of Figure 5. 
In summary, a final constraint on log\,$U$ that
allows for the possibility that the doubly-ionized species
in the $-$300\kms\ component are collisionally ionized 
(and hence should not be included in the CLOUDY model)
is between $-$3.45 and $-$3.15. 

The range of allowed values for log\,$U$ allows us to constrain
the physical conditions in the MS toward \qb.
We derive a gas density log\,($n_{\rm H}$/cm$^{-3}$) of
$-$2.51 to $-$2.21, 
a neutral hydrogen fraction of 0.5--1.1\%
(ionization fraction H$^+$/H of 98.9--99.5\%),
a total hydrogen column density log\,$N$(\hi+\hw) of 18.64--18.96,
a line-of-sight size $l$ of 0.2--0.9\,kpc, 
and a gas pressure $P$/k=2.1$nT$ of 90--170\,cm$^{-3}$\,K 
(the factor of 2.1 accounts for the electron pressure).
Repeating the model with the EGB radiation field only (no MW field)
only changes the derived value of log\,$U$ by 0.05\,dex. 

Whereas the best-fit model is capable (by design) of
explaining the low-ion (singly- and doubly-ionized species) column
densities, it cannot explain the high ions (triply-ionized and above).
The log\,$U$=$-$3.45 model under-produces the actual \sif, \cf, and
\os\ columns by  1.6, 2.5, and $>$8 orders of magnitude, respectively.
This is shown for \cf\ and \sif\ in Figure 5b 
(compare the models and data points for the blue and green diamonds). 
The difficulty of photoionizing the high ions in HVCs has been found
before by many authors 
\citep{Se03, Tr03, GS04, Co04, Co05, Co07, Fo05a, Ga05}, and indicates
a separate ionization mechanism is required for the high ions.
The separate behavior of the high ions is confirmed by 
their different (but overlapping) component structure in velocity space.

Toward \qc, an AGN sightline 11\dgr\ Off-Stream
(i.e., off the 21\,cm-emitting regions) where the MS is seen in
absorption with $N$(\hi)$\approx$17.0, \citet{Fo05a} find CLOUDY solutions 
to the \sit/\siw\ and \ct/\cw\ ratios with log\,$U$ between $-$3.5 and
$-$3.7, implying hydrogen ionization fractions $>$98\%. 
Our new results for the \qb\ sight line are in reasonable agreement with
this for the ionization level of the low-ion gas in the outer MS (the
Off-Stream directions).

\section{High Ions: Collisional Ionization Modeling}

\begin{figure*}[!ht]
\epsscale{1.0}\plotone{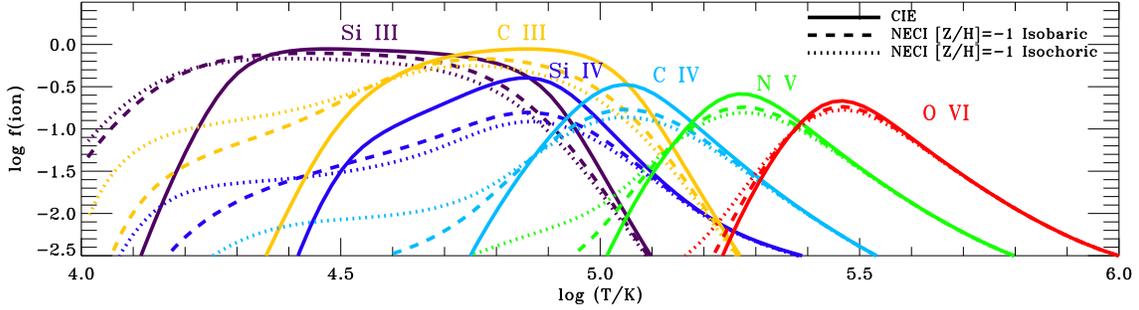}
\caption{Ionization fractions versus temperature in the \citet{GS07}
  collisional ionization equilibrium (CIE) and non-equilibrium collisional
  ionization (NECI) models, for \sit, \ct, and the four high ions
  \sif, \cf, \nf, and \os. }
\end{figure*}

\begin{figure}[!ht]
\epsscale{1.2} 
\plotone{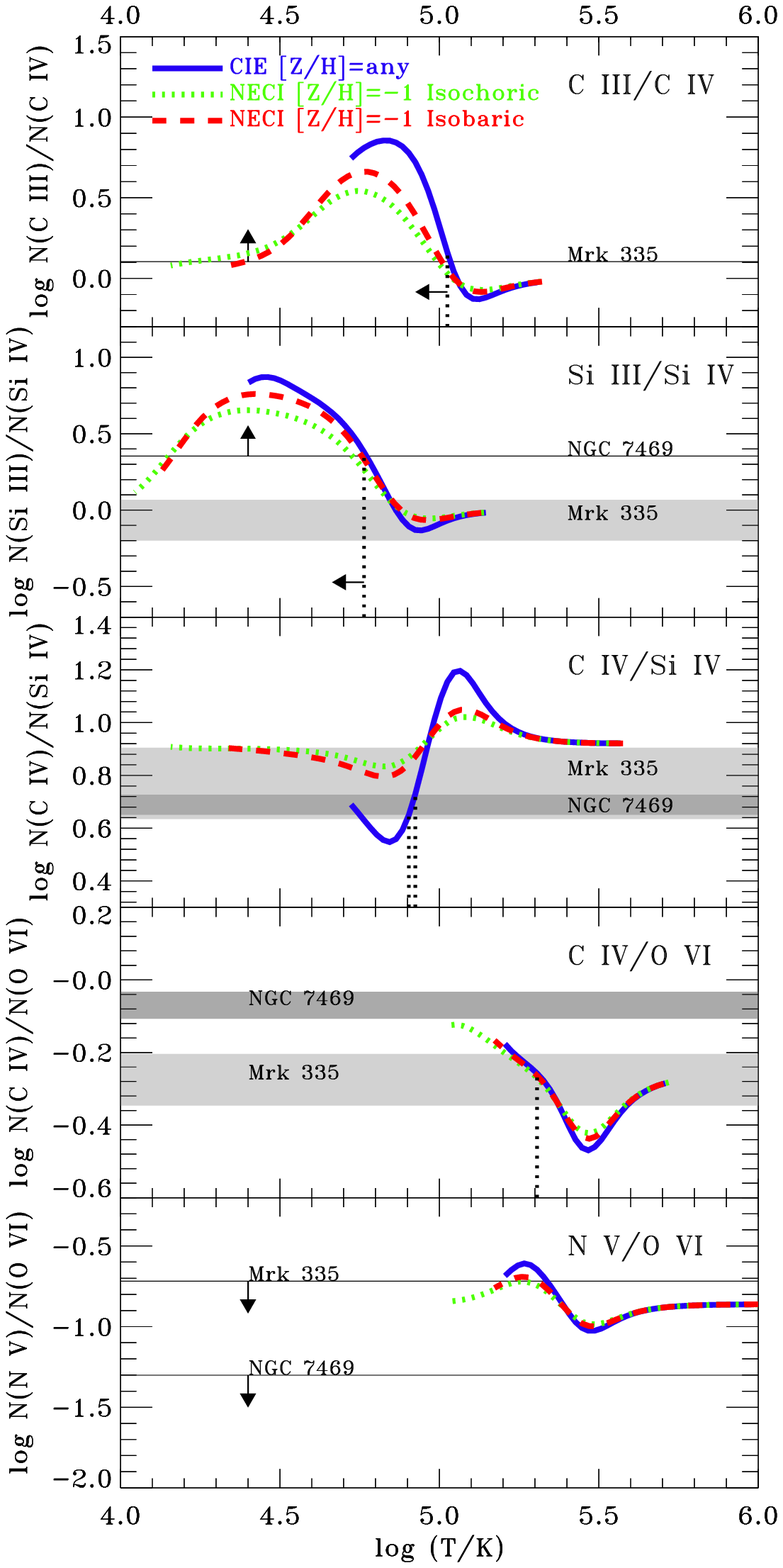}
\caption{Collisional ionization (CI) modelling of the high-
  and intermediate ions in the MS. 
  Each panel displays a different column density ratio.
  The colored lines show the predicted ratios
  versus temperature in the CIE and NECI models of \citet{GS07}
  assuming solar relative abundances. 
  The observed ratios (calculated using the AOD column densities in
  the MS components) are shown as shaded regions, or arrows in the
  case of upper/lower limits.
  The best-fit values (and limits on) of log\,$T$ for the CIE case are
  shown with vertical dotted lines.
  The model predictions are only displayed for temperatures where the
  ion fraction is $>$0.001 for both ions (see text).}
\end{figure}

As photoionization is unable to explain the observed strength of the
high-ion absorption lines in the MS (or in other HVCs), 
we turn to collisional ionization (CI). We use the CI models of
\citet[][hereafter GS07]{GS07}, which predict the
ionization fractions as a function of temperature for all commonly observed
elements, in both the equilibrium (CIE) and non-equilibrium (NECI) cases. 
Figure 6 shows these predicted ionization fractions 
for \sit, \ct\ and the four high ions \sif, \cf,
\nf, and \os. We show the CIE case and two NECI models at [Z/H]=$-$1 
(the closest model to the MS metallicity), for the isobaric and
isochoric cases.
Other models have been developed to explain the high-ion plasma
around interstellar clouds, including
conductive interfaces \citep{BH87, Bo90, GS10}, 
turbulent mixing layers \citep{Sl93, Es06}, and shock ionization
\citep{DS96, IS04}. Summaries of the predicted high-ion ratios in
many of these models are given in \citet{Sp96} and \citet{Fo04}.

To compare the CI models with the MS observations, we analyze the 
high-ion column density ratios (Figure 7).
The main panels display the run of five ion ratios 
(\ct/\sif, \sit/\sif, \cf/\sif, \cf/\os, and \nf/\os) vs temperature 
in the CIE and NECI models.
The latter three of these ratios assume solar relative abundances.
The shaded regions show the observed high-ion ratios in the MS
component, using the AOD column densities reported in Tables 1 and
2. In principle, the high-ion $b$-values can be used to give an upper
limit on the temperature, but the observed limits are not constraining:
for example, the measured $b$(\cf)=28\kms\ in the MS component toward
\qb\ implies log\,$T\!<\!5.75$ in the \cf\ phase. 
The model high-ion ratios are only plotted for temperatures where both
ion fractions $f$ are greater than some threshold 
(here arbitrarily chosen as 0.001), otherwise the total column 
density, which scales as $f^{-1}$, becomes implausibly large.
For example, the \cf/\os\ ratio is only plotted at temperatures where
\cf/C$>$0.001 and \os/O$>$0.001. 

If the observed high-ion lines in the MS were formed in a single region of
plasma in CIE or NECI, all the ratios on Figure 7 would yield a solution at
the same temperature. This is clearly not the case;
\emph{in both sight lines, there is no single-temperature CI solution
  (in or out of equilibrium) that can explain all three of the \sif,
  \cf, and \os\ column densities.} 
This suggests that the MS high-ion plasma is itself multi-phase,
with one phase containing the \sif, and one containing the \os, and
the \cf\ potentially contributing to each. We therefore look at each
ratio in turn to investigate multi-phase solutions. 

The top two panels of Figure 7 show the observed \ct/\cf\ and
\sit/\sif\ ratios in 
the MS. The model predictions for these ratios have the advantage of
not depending on the elemental abundances. 
The \ct/\cf\ ratio toward \qb\ ($>$0.43 in the log) 
can be reproduced by gas in CIE at 4.75$<$log\,$T\!<$4.96. 
The \sit/\sif\ ratio toward \qa\ ($>$0.49 in the log) 
can be reproduced by gas in CIE at 4.40$<$log\,$T\!<$4.72, and 
toward \qb, the \sit/\sif\ ratio (0.22$\pm$0.13 in the log) 
can be explained by CIE at log\,$T$=4.80$\pm$0.03. 

The \cf/\sif\ ratio toward \qa\ (0.69$\pm$0.04 in the log)
can be reproduced with a CIE model at log\,$T$=4.90$\pm$0.02 if [C/Si]=0.
This model predicts $N_{\rm a}$(\sit)=13.91, to be
compared with the observed $N$(\sit)$>$13.82 in the MS.
That is, a log\,$T$=4.90 CIE model can explain the \cf, \sit, and
\sif\ MS column densities toward \qa. 
The \cf/\sif\ ratio toward \qb\ (0.77$\pm$0.13 in the log)
is also reproduced by a CIE model at log\,$T$=4.92$\pm$0.03 if 
[C/Si]=0.
This CIE model predicts $N$(\ct)=14.29 and $N$(\sit)=13.44,
to be compared with the observed $N_{\rm a}$(\ct)$>$14.00 
and $N_{\rm a}$(\sit)=13.01, 
i.e. it over-predicts the \sit\ column by 0.43\,dex.
On a linear scale, our measured \cf/\sif\ ratios are 
4.9$^{+0.4}_{-0.4}$ toward \qa\ and 5.9$^{+2.1}_{-1.6}$
toward \qb, fairly close to the Galactic halo average of
3.5$\pm$1.1 reported by \citet{Zs03}, but slightly higher than the
Magellanic Bridge values of between 
1.5$^{+1.5}_{-0.8}$ and 2.3$^{+1.6}_{-1.1}$ reported by \citet{Le02}.

Note that in \S6, we included the \ct\ and \sit\ ions in the
photoionization models, i.e. we placed them in the \emph{low-ion} phase
of the MS. In both sightlines, this approach is justified in the range
$-$360 to $-$320\kms\ where \cw, \ct, \siw, and \sit\ all show a
similar component structure. 
However, the twin findings that (a) in both sightlines, \sit\ shows
absorption in the $-$300\kms\ high-ion component seen in \cf\ and \os,
and (b) a GS07 CIE model to the measured \cf/\sif\ ratio in the
$-$300\kms\ component 
predicts fairly successfully the observed \ct\ and \sit\ columns in
that component, indicate that \emph{collisional ionization 
cannot be ignored as a contributor to the doubly-ionized species in the MS}.
Hybrid photo+collisional-ionization models are needed to properly
account for this \citep[see][]{Tr08,Ho09}. 

We can also investigate solutions that apply if the \cf\ and
\os\ are co-spatial.
The observed MS \cf/\os\ ratios 
have no CIE or NECI solution unless the C/O ratio is super-solar: 
toward \qa\ where \cf/\os\ = $-$0.07$\pm$0.04 (logarithmic), 
a CIE solution is possible at log\,$T$=5.25$\pm$0.04 if 
[C/O]=+0.15\footnotemark[8]\footnotetext[8]{This ratio seems 
  unlikely, since low-metallicity environments (e.g. metal-poor halo
  stars) show a tendency for sub-solar C/O ratios \citep{Ak04}.}. 
Toward \qb, the logarithmic \cf/\os\ ratio of $-$0.27$\pm$0.07 
has a CIE solution at log\,$T$=5.25$\pm$0.04 if [C/O]=0. 
Toward \qa, no \nf\ is detected in the MS component
(although \nf\ is detected in HVC2);
we place a limit log\,$N$(\nf)/$N$(\os)$<-$1.30 in the MS component,
requiring a sub-solar [N/O]$_{\rm MS}$ (as in the SMC) for 
a CI solution to be be found. The non-detection of \nf\ toward \qb\
is not strong enough to constrain [N/O].

\subsection{Total Hydrogen Column Density in the High-Ion Phases}
Although the ambiguity over the location of the \cf\
prevents us from drawing firm conclusions  
about the temperature in the high-ion regions, we can still infer
lower limits on the total column density of hot plasma in the MS
from the \sif\ and \os\ observations alone.
This is because the \sif\ and \os\ ion fractions \sif/Si and \os/O
never exceed certain maximal values (Figure 6).
It can be seen that \sif/Si$<$0.40 and \os/O$<$0.22 in \emph{any}
of the GS07 CI models (CIE or NECI).
Thus, the \hw\ column in the \sif\ phase can be calculated as
$N$(\hw)$_{\rm Si~IV}\!>\!N$(\sif)/[(\sif/Si)$_{\rm max}$(Si/H)],
and the \hw\ column in the (separate) \os\ phase as
$N$(\hw)$_{\rm O~VI}\!>\!N$(\os)/[(\os/O)$_{\rm max}$(O/H)], where we assume 
that [Si/H] and [O/H] in the high-ion phase equal the metallicity
measured in the low-ion phase, [O/H]$_{\rm MS}$=$-$1.00. 
If [Si/O]$<$0 in the high-ion phase, as is the case in the low-ion phase, 
then the value of $N$(\hw)$_{\rm Si~IV}$ increases accordingly. Hence
the lower limit on $N$(\hw)$_{\rm Si~IV}$ is doubly conservative, once
for the ionization correction and once for the abundance correction.

In the MS toward \qa\ where 
log\,$N_{\rm a}$(\sif)=13.33 and log\,$N_{\rm a}$(\os)=14.09, we derive 
$N$(\hw)$_{\rm Si~IV}\!>\!19.2$ and $N$(\hw)$_{\rm O~VI}\!>\!19.1$. 
In the MS toward \qb\ 
with log\,$N_{\rm a}$(\sif)=12.80 and log\,$N_{\rm a}$(\os)=13.84, we derive
$N$(\hw)$_{\rm Si~IV}\!>\!18.7$ and $N$(\hw)$_{\rm O~VI}\!>\!18.8$ 
We emphasize that the \hw\ columns derived here are distinct from
the warm \hw\ columns derived in the photoionized phase of the MS in
\S6, since the high ions trace physically different regions.
The total plasma content in the MS is then\\
$N$(total \hw)=
$N$(\hw)$_{\rm warm}$+$N$(\hw)$_{\rm Si~IV}$+$N$(\hw)$_{\rm O~VI}$.

\section{Ionized-to-Neutral Ratio in the Stream}
By summing the lower limits on $N$(\hw) in the low-ion and
high-ion phases of the MS as calculated in \S6 and \S7, we find 
$N$(total \hw)$\ga$19.9 toward \qa\ and $\ga$19.2 toward \qb. %
The ratio $N$(total \hw)/$N$(\hi), indicating the relative amounts of
ionized and neutral gas in the MS, is then 
$\ga$19 toward \qa, and $\ga$330 toward \qb. 
Along the sight line toward \qc, 11\dgr\ away from the MS,
we use the column densities presented by \citet{Fo05a} to derive
$N$(total \hw)/$N$(\hi)$\ga$120 using an identical method. 
These numbers, summarized in Table 4, indicate that along each of
these three sight lines, the MS contains significantly more 
ionized gas than neutral gas. 

The detection of high-ion absorption together with the observed
fragmentation of the Stream's \hi\ into small clumps \citep{St02,
  St08, Wa02, WK08, Ni09} supports the picture that
the MS is being ``evaporated'' as it plows through the hot Galactic
corona, a scenario predicted by hydrodynamical simulations
\citep{HP09, BH09} and theoretical considerations 
\citep[the dominance of heating over drag;][]{Mu00}. 
Indeed, the maps presented by \citet{St08} show that the MS filament ``S3''
breaks into clumps very close to the \qa\ direction under study here.
Evaporation has important consequences for the role HVCs play in Galactic
evolution, since if they cannot survive their inflow passage, they
will serve to replenish the hot corona with new plasma, rather than to bring
neutral fuel to the disk for future generations of star formation. 
However, the leading arm of the MS has apparently reached the Galactic
disk where it is interacting with the neutral gas \citep{MG08}, so
a stabilizing mechanism \citep[e.g., magnetic fields;][]{Ko02} 
may protect it against disruption. 

If the MS and other HVCs break up as they
move through the hot halo, one may expect to see short-lived
circumgalactic gas fragments as a result.
Indeed, a recent survey of high-velocity \oi\ absorption
along 26 QSO sightlines uncovered a previously unnoticed population
of low-column density circumgalactic clouds \citep{Ri09}.
These low-mass clouds represent Lyman-Limit systems and thus
remain mostly unseen in 21\,cm emission, but cover a significant
fraction ($\sim$25\%) of the sky and must be very numerous 
($\ga10^8$) and widespread throughout the Milky Way halo. 
The sky position and radial velocities of some of the detected
\oi\ features are consistent with a MS origin, supporting the
idea that the MS has already lost a
substantial portion of its original mass during its passage through
the halo, leaving behind debris that is rapidly
evaporated and incorporated into the Milky Way's coronal gas
reservoir. This debris may also be seen via optical observations of
high-velocity \caw\ absorption \citep{Ri05, Sm05, Be08}.

\begin{deluxetable*}{lcccc ccc}
\tablewidth{0pt}
\tabcolsep=4.0pt
\tabletypesize{\small}
\tablecaption{Hydrogen Column Densities in the Gaseous Phases of the MS}
\tablehead{Sightline & Location & $N$(\hi)\tm{a} & \multicolumn{4}{c}{\underline{~~~~~~~~~~~~~~~~~~$N$(\hw)~~~~~~~~~~~~~~~~~~}} & $\frac{N({\rm Total~H~II})}{N({\rm H~I})}$\\  & & & Low-ion\tm{b} & \sif\tm{c} & \os\tm{c} & Total & }
\startdata
  NGC 7469 &   On Stream &  18.63$\pm$0.08       & $>$19.8 & $\ga$19.2 & $\ga$19.1 & $\ga$19.9 & $\ga$ 19 \\
   Mrk 335 &  Off Stream &  16.67$\pm$0.05       & $>$18.6 & $\ga$18.7 & $\ga$18.8 & $\ga$19.2 & $\ga$330 \\
                             HE~0226-4110\tm{d} & Off-Stream & $\approx$17.0 & $\approx$19.0   & $\ga$18.7 & $\ga$18.9 & $\ga$19.4 & $\ga$220
\enddata
\tablecomments{The measured MS metallicity [O/H]=$-$1.0 is assumed to apply in all phases.}
\tn{a}{For \qa, $N$(\hi) is measured from the GBT 21\,cm spectrum. For the other two sightlines, $N$(\hi) is measured using the higher-order Lyman lines in the \fuse\ spectra.\\}
\tn{b}{$N$(\hw) in low-ion phase calculated using CLOUDY models described in \S5.1.\\}
\tn{c}{~Lower limits on $N$(\hw) in \sif\ and \os\ phases calculated using the maximal collisional ionization fractions described in \S5.2.\\}
\tn{d}{Using results from \citet{Fo05a}, integrated over \vlsr=80--230\kms, with low-ion ionization level 99\%.}
\end{deluxetable*}

\section{Summary}
We have used \fuse\ and \hst/STIS/E140M UV spectra of two AGN
to study the Magellanic Stream in absorption, focusing on its
ionization level and chemical enrichment.
The AGN are \qa, lying behind the MS in a direction with 
log\,$N$(\hi)$_{\rm MS}$=18.63$\pm$0.03(stat)$\pm$0.08(syst) measured
from GBT and Effelsberg data, and \qb\ lying 24.9\dgr\ away with 
log\,$N$(\hi)$_{\rm MS}$=16.67$\pm$0.05 measured from \hi\ Lyman series
absorption. 
These data are supplemented by VLT/UVES optical absorption profiles of 
\caw\ and \nao\ toward \qa. Our results can be summarized
as follows. 

\begin{enumerate}
\item In both sight lines, the MS is detected as an absorption-line
  HVC at $v_{\rm LSR}\!\approx\!-$340\kms.
  Other HVCs observed in both directions within $\sim$100\kms\ of the
  main component may also be related to the MS.
  The detected species include the low ions 
  \hi, \oi, \cw, \ct, \siw, \sit, \few, \fet, \alw, and \caw,
  and the high ions \cf, \sif, and \os.
  These lines are all weaker in the \qb\ (Off-Stream) direction. 
  The detection of the MS in UV absorption in 
  two Off-Stream directions (\qb\ and \qc) demonstrates that
  even the most sensitive 21\,cm observations do not reveal all the
  gas in the MS.

\item Combining our UV measurement of $N$(\oi)
  in the MS toward \qa\ with $N$(\hi) measured from the 21\,cm data, 
  we derive a metallicity 
  [O/H]$_{\rm MS}$=$-$1.00$\pm$0.05(stat)$\pm$0.08(syst)
  (10$\pm$1$\pm$2\% of solar), 
  where the systematic error accounts for the beam-size mismatch
  between the radio and UV observations.
  This value, which is robust against ionization corrections, is 
  similar to the Magellanic Bridge oxygen abundance measured by
  \citet{Le08}, and  
  is closer to the SMC abundance than the LMC abundance, 
  supporting the view that the Stream originates in the SMC.
  We also measure [N/H]$_{\rm MS}\!<\!-0.44$ (3$\sigma$). 

\item A multi-phase structure to the MS, with a warm, photoionized
  low-ion phase and hotter, collisionally-ionized high-ion phase(s),
  is required by two independent findings: 
  (1) the high-ion profiles differ from the low-ion profiles, with the
  high ions showing an additional component at $-$300\kms\ not seen in
  the low ions, and
  (2) a single-phase photoionization model cannot
  simultaneously reproduce both the low-ion and high-ion column
  densities in the MS.

\item We investigated the ionization level in the \emph{low-ion} phase
  of the MS with the photoionization code CLOUDY, using a radiation
  field that includes both Galactic and extragalactic contributions. 
  The model assumes that the \hi\ and the singly- and doubly-ionized
  species are co-spatial.
  Toward \qa, our limit on the \sit/\siw\ ratio allows us
  to constrain the ionization parameter log\,$U\!>\!-$3.30, giving 
  a MS gas density log\,($n_{\rm H}$/cm$^{-3}$)$<\!-$2.36, 
  a neutral fraction $x_{\rm H}\!<\!0.07$ 
  (hydrogen ionization level H$^+$/H$>$93\%), 
  a total low-ion hydrogen column density 
  log\,$N$(\hi+\hw)$>$19.79,\ 
  and a line-of-sight cloud size $l\!>$4.6\,kpc.
  The model finds [Si/O]$_{\rm MS}\!<$--0.3, [Fe/O]$_{\rm MS}\!<$--0.2, and 
  [Al/O]$_{\rm MS}\!<$--0.7; these sub-solar abundances provide evidence
  for depletion onto dust grains. 
  Toward \qb, we use the \sit/\siw\ and \ct/\cw\ ratios to constrain
  log\,$U$ to lie between $-$3.45 to $-$3.15, giving 
  log\,($n_{\rm H}$/cm$^{-3}$) between $-$2.51 and $-$2.21,
  $x_{\rm H}$=0.5--1.1\% (H$^+$/H=98.9--99.5\%),
  log\,$N$(\hi+\hw)=18.64--18.96,
  $l$=0.2--0.9\,kpc, 
  and a gas pressure $P$/k=90--170\,cm$^{-3}$\,K. 
  The range in the above values accounts for the possibility of a
  collisionally-ionized contribution to the \sit\ and \ct\ columns
  from the $-$300\kms\ component.
  
\item The \emph{high-ion} column densities in the MS are significantly
  underestimated by photoionization models, so collisional ionization is
  required. This process can occur at the interfaces between the warm
  phase of the Stream and the hot Galactic corona. 
  However, there is no single-temperature CI solution that explains
  \cf, \sif, and \os, suggesting multiple high-ion phases are present:
  one seen in \sif, one seen in \os, with \cf\ potentially
  contributing to each.  
  The \cf/\sif\ ratios of 4.9$^{+0.4}_{-0.4}$ toward \qa\ 
  and 5.9$^{+2.1}_{-1.6}$ toward \qb\ 
  can be explained by plasma in CIE at log\,$T$=4.90$\pm$0.02
  and 4.92$\pm$0.03 respectively, assuming [C/Si]=0. 
  Such models can contribute to the \ct\ and \sit\ columns. 
  Assuming that the metallicity measured in the
  low-ion phase applies to the entire MS complex, we infer that the
  H$^+$ column in the \sif\ and \os\ phases is substantial: 
  log\,$N$(\hw)$_{\rm Si~IV}\!\ga\!18.7$ and 
  log\,$N$(\hw)$_{\rm O~VI}\!\ga\!18.8$ in both sightlines. 

\item Summing over the low-ion and both high-ion phases, 
  conservative lower limits on the total \hw\ columns in the MS are 
  log\,$N$(total \hw)$\ga$19.9 toward \qa, and $\ga$19.2 toward \qa. 
  Therefore, the total-ionized-to-neutral ratio 
  $N$(total \hw)/$N$(\hi) in the MS is $\ga$19 toward \qa, $\ga$330 
  toward \qb, and $\ga$220 toward \qc, 
  a third MS sight-line with results taken from the literature.
  \emph{That is, along these three sight lines, the vast majority
  ($>$95\%) of the Stream has been ionized}.
  The presence of warm-hot plasma around the MS (seen in the high ions)
  supports the picture that the Stream is being ``evaporated'' as it
  plows through the hot Galactic corona and will not reach the disk,
  as predicted by hydrodynamical simulations and suggested by the
  observed fragmentation of the Stream's \hi\ into small clumps.
\end{enumerate}

It remains to be seen whether the MS represents a fundamentally different
category of object than the other large HVCs (Complex C, Complex A, Cohen
Stream, Smith Cloud) whose distances are all
within a factor of two of 10\,kpc \citep{Wa07, Wa08, Th08}.
There is an interesting contradiction in that the MS, out at
$d\!\sim\!50$\,kpc, is apparently in the process of being destroyed by its
interaction with the corona, but yet the nearby HVCs at $d\!\sim\!10$\,kpc 
have managed to reach their current position without breaking apart.
This paradox may be solved if HVCs are short-lived structures, constantly
condensing out of, and re-evaporating into, the hot halo,
or alternatively, if they are stabilized against evaporation.
Further UV observations with the Cosmic Origins Spectrograph (COS),
now installed on \hst, will be invaluable for constraining the
MS ionization level in directions covering a wide range of $N$(\hi).
Such measurements are needed to determine 
the Galactic accretion rate in ionized gas, and the role of HVCs
in the global process of Galactic metabolism.\\

{\it Acknowledgments}.
The \fuse\ satellite retired from service in October 2007.
We thank the \fuse\ team at Johns Hopkins University for their
tireless efforts to develop, operate, and support this productive facility.
We thank Marilyn Meade for expert assistance with the \fuse\ data
reduction pipeline, and the referee for a useful report.
AJF gratefully acknowledges support from an ESO Fellowship.
Support for program GO-9802 was provided by NASA through a grant from
the Space Telescope Science Institute, which is operated by the
Association of Universities for Research in Astronomy, Inc., under
NASA contract NAS 5-26555.

\end{document}